\documentstyle[referee,psfig,epsf]{l-aa}
\def\lsim{\lower.5ex\hbox{$\; \buildrel < \over \sim \;$}}
\def\gsim{\lower.5ex\hbox{$\; \buildrel > \over \sim \;$}}
\newcommand{\eqb}{\begin{eqnarray}}
\newcommand{\eqe}{\end{eqnarray}}

\newbox\grsign \setbox\grsign=\hbox{$>$} \newdimen\grdimen \grdimen=\ht\grsign
\newbox\simlessbox \newbox\simgreatbox
\setbox\simgreatbox=\hbox{\raise.5ex\hbox{$>$}\llap
   {\lower.5ex\hbox{$\sim$}}}\ht1=\grdimen\dp1=0pt
\setbox\simlessbox=\hbox{\raise.5ex\hbox{$<$}\llap
   {\lower.5ex\hbox{$\sim$}}}\ht2=\grdimen\dp2=0pt
\def\gsim{\mathrel{\copy\simgreatbox}}
\def\lsim{\mathrel{\copy\simlessbox}}

\def\ref{\par\noindent\hangindent=1.5cm}
\begin{document}
\thesaurus{06(02.01.2, 02.02.1, 08.14.1, 02.19.1, 02.14.1)}
\title{Nucleosynthesis in Accretion Flows Around Black Holes}
\author{Banibrata Mukhopadhyay and Sandip K. Chakrabarti} 
\institute{S.N. Bose National Centre for Basic Sciences,
JD Block, Salt Lake, Sector-III, Calcutta-700091, INDIA}

\offprints{Banibrata Mukhopadhyay}
\date{Astromomy \& Astrophysics (in press) }
\maketitle
\markboth{Nucleosynthesis in Disks}{}

\maketitle


\begin{abstract}

Significant nucleosynthesis is possible in the centrifugal pressure-supported dense and hot region of the accretion flows 
which deviate from Keplerian disks around black holes. 
We compute composition changes and energy generations 
due to such nuclear processes. We use a network 
containing $255$ species and follow the changes in composition. 
Highly viscous, high-accretion-rate flows deviate from a 
Keplerian disk very close to the black hole and the 
temperature of the flow is very small due to Compton cooling. 
No significant nucleosynthesis takes place in these cases. Low-viscosity
and lower-accretion-rate hot flows deviate farther out and
significant changes in composition are possible in these
cases. We suggest that such changes in composition could 
be contributing to the metallicities of the galaxies. 
Moreover, the radial variation of the energy 
generation/absorption specifically due to proton capture 
and photo-dissociation reactions could cause 
instabilities in the inner regions of the
accretion flows. For most of these cases sonic
point oscillations may take place. We discuss
the possibility of neutrino emissions.

\end{abstract}

\section{Introduction}

In Chakrabarti \& Mukhopadhyay (1999, hereafter referred to as Paper 1)
we studied the result of nucleosynthesis in hot, highly viscous  accretion flows with
small accretion rates and showed that neutron tori can form around a 
black hole. In the present paper, we study nucleosynthesis in disks in other
parameter space, where the photo-dissociation may not be complete and other
reactions may be important, and show that depending on the
accretion parameters, abundances of new isotopes may become abnormal
around a black hole. Thus, observation of these isotopes may give a possible 
indication of black holes at the galactic center or in a binary system.

Earlier, Chakrabarti (1986) and Chakrabarti et al.(1987, hereinafter
CJA) initiated discussions of nucleosynthesis in sub-Keplerian disks around 
black holes and concluded that for very low viscosity ($\alpha$ parameter
less than around $10^{-4}$) and high accretion rates (typically,
ten times the Eddington rate) there could be significant nucleosynthesis
in thick disks. Radiation-pressure-supported thick accretion flows are cooler
and significant nucleosynthesis was not possible unless the residence
time of matter inside the accretion disk was made sufficiently high
by reducing viscosity. The conclusions of this work were later verified
by Arai \& Hashimoto (1992) and Hashimoto et al. (1993).

However, the theory of accretion flows which contain a centrifugal-pressure-supported hotter and denser region in the inner part of the accretion disk 
has been developed more recently (Chakrabarti 1990, hereafter C90
and Chakrabarti 1996, hereafter C96). The improvement in the
theoretical understanding can be appreciated by comparing the 
numerical simulation results done in the eighties
(e.g. Hawley et al. 1984, 1985) and in the nineties
(e.g.  Molteni et al. 1994; Molteni et al.
1996; Ryu et al. 1997). 
Whereas in the eighties the matching of theory and numerical
simulations was poor, the matching of the
results obtained recently is close to perfect. It is realized
that in a large region of the parameter space, especially for lower
accretion rates, the deviated flow would be hot and a significant 
nuclear reaction is possible without taking resort to very low viscosity.

We arrive at a number of the important conclusions: (a) Significant nucleosynthesis
is possible in the accretion flows. Whereas most of the matter of modified
composition enters inside the black hole, a fraction may go out
through the winds and will contaminate the surroundings
in due course. The metalicity of the galaxies may also be influenced.
(b) Generation or absorption of energy due to exothermic 
and endothermic nuclear reactions could seriously affect the stability of a disk. 
(c) Hot matter is unable to produce Lithium ($^7\!Li$) or Deuterium (D) since
when the flow is hot, photo-dissociation (photons partially locally generated 
and the rest supplied by the nearby Keplerian disk (Shakura \& Sunyaev 1973) when 
the region is optically thin) is enough to dissociate all the elements 
completely into protons and neutrons. Even when photo-dissociation is turned off
(low opacity cases or when the system is fundamentally photon-starved) 
$Li$ was not found to be produced very much.
(d) Most significantly, we show that one does not require a 
very low viscosity for nucleosynthesis in contrary to the 
conclusions of the earlier works in thick accretion disk (e.g., CJA).

In Paper 1, we already presented the basic equations which 
govern accretion flows around a compact object, so we do not 
present them here. The plan of the present paper is the following: 
we present a set of solutions of these equations in the next section 
which would be used for nucleosynthesis work. 
When nucleosynthesis is insignificant, we compute 
thermodynamic quantities ignoring nuclear energy generation, otherwise we
include it. The detailed method is presented here. We divide all the disks 
into three categories: ultra-hot, moderately hot, and cold. In Sect. 3, 
we present the results of nucleosynthesis for these cases. We find 
that in ultra-hot cases, the matter is completely photo-dissociated. 
In moderately hot cases, proton-capture processes 
along with dissociation of deuterium
and $^3\!He$ are the major processes. In the cold cases, no
significant nuclear reactions go on. In Sect. 4, we discuss the stability properties of the
accretion disks in presence of nucleosynthesis and conclude that only the very 
inner edge of the flow is affected. Nucleosynthesis
may affect the metallicities of the galaxies as well as $Li$ abundance
in companions in black hole binaries. In Sect. 5, we discuss these issues and draw our conclusions.

\section{Typical Solutions of Accretion Flows}

In our work below, we choose a Schwarzschild black hole and use the Schwarzschild 
radius $2GM/c^2$ to be the unit of the length scale where $G$ and $c$ are the gravitational constant 
and the velocity of light respectively. We choose 
$c$ to be the unit of velocity. We also choose the  cgs unit when we find it
convenient to do so. The nucleosynthesis work is done
using cgs units and the energy release rates are in that unit as well.

A black hole accretion disk must, by definition, have {\it radial} motion, 
and it must also be transonic, i.e., matter must be supersonic (C90) while
entering through the horizon.  The supersonic flow must be sub-Keplerian
and therefore deviate from the Keplerian disk away from the
black hole. The location where the flow may deviate will depend
on the cooling and heating processes (which depend on viscosity). Several solutions
of the governing equations (see Eq. 2(a-d) of Paper 1) are given in C96. 
By and large, we follow this paper to compute thermodynamical 
parameters along a flow. However, we have considered Comptonization
as in Chakrabarti \& Titarchuk (1995, hereafter CT95) and Chakrabarti (1997, hereafter C97).
Due to computational constraints, we include energy generation due to nuclear 
reactions ($Q_{\mathrm{nuc}}$) only when it is necessary (namely, when
$|Q_{\mathrm{nuc}}|$ is comparable to energy generation due to viscous 
effects), and we do not consider energy generation due to magnetic
dissipation (due to reconnection effects, for instance). 
In Fig. 1, we show a series of solutions which we 
employ to study nucleosynthesis processes. We plot the ratio 
$\lambda/\lambda_K$ (Here, $\lambda$ and $\lambda_K$ are the specific
angular momentum of the disk and the Keplerian angular momentum respectively.)
as a function of the logarithmic radial distance.
The coefficient of the viscosity parameters are marked on each curve.
The other parameters of the solution are in Table 1. 
These solutions are obtained with constant $f=1-Q^-/Q^+$ and 
$Q^+$ include only the viscous heating. In presence of significant nucleosynthesis,
the solutions are obtained by choosing $f=1-Q^-/(Q^++Q_{\mathrm{nuc}})$, where $Q_{\mathrm{nuc}}$ is the
net energy generation or absorption due to exothermic and endothermic reactions. 
The motivation for choosing the particular cases are mentioned in the next section.
At $x=x_{K}$, the ratio $\lambda/\lambda_K=1$ and therefore $x_{K}$
represents the transition region where the flow deviates from a Keplerian disk.
First, note that when other parameters (basically, specific angular momentum
and the location of the inner sonic point) remain roughly the same,
$x_{K}$ changes inversely with viscosity parameter $\alpha_\Pi$ (C96). 
(The only exception is the curve marked with $0.01$. This is because it is drawn 
for $\gamma=5/3$; all other curves are for $\gamma=4/3$.) If one assumes, 
as Chakrabarti \& Titarchuk (1995) and Chakrabarti (1997) 
did, that the alpha viscosity parameter {\it decreases}
with vertical height, then it is clear from the general behaviour of Fig. 1
that $x_K$ would go up with height. The disk will then look like a sandwich
with higher viscosity Keplerian matter flowing along the equatorial plane. 
As the viscosity changes, the sub-Keplerian and Keplerian flows redistribute
(Chakrabarti \& Molteni 1995) and the inner edge of the Keplerian component
also recedes or advances. This fact that the inner edge of the disk
should move in and out when the black hole goes into soft or hard state 
(as observed by, e.g., Gilfanov et al. 1997; Zhang et al. 1997) is
thus naturally established from this disk solution.

\begin {figure}
\vbox{
\vskip +0.0cm
\hskip 0.0cm
\centerline{
\psfig{figure=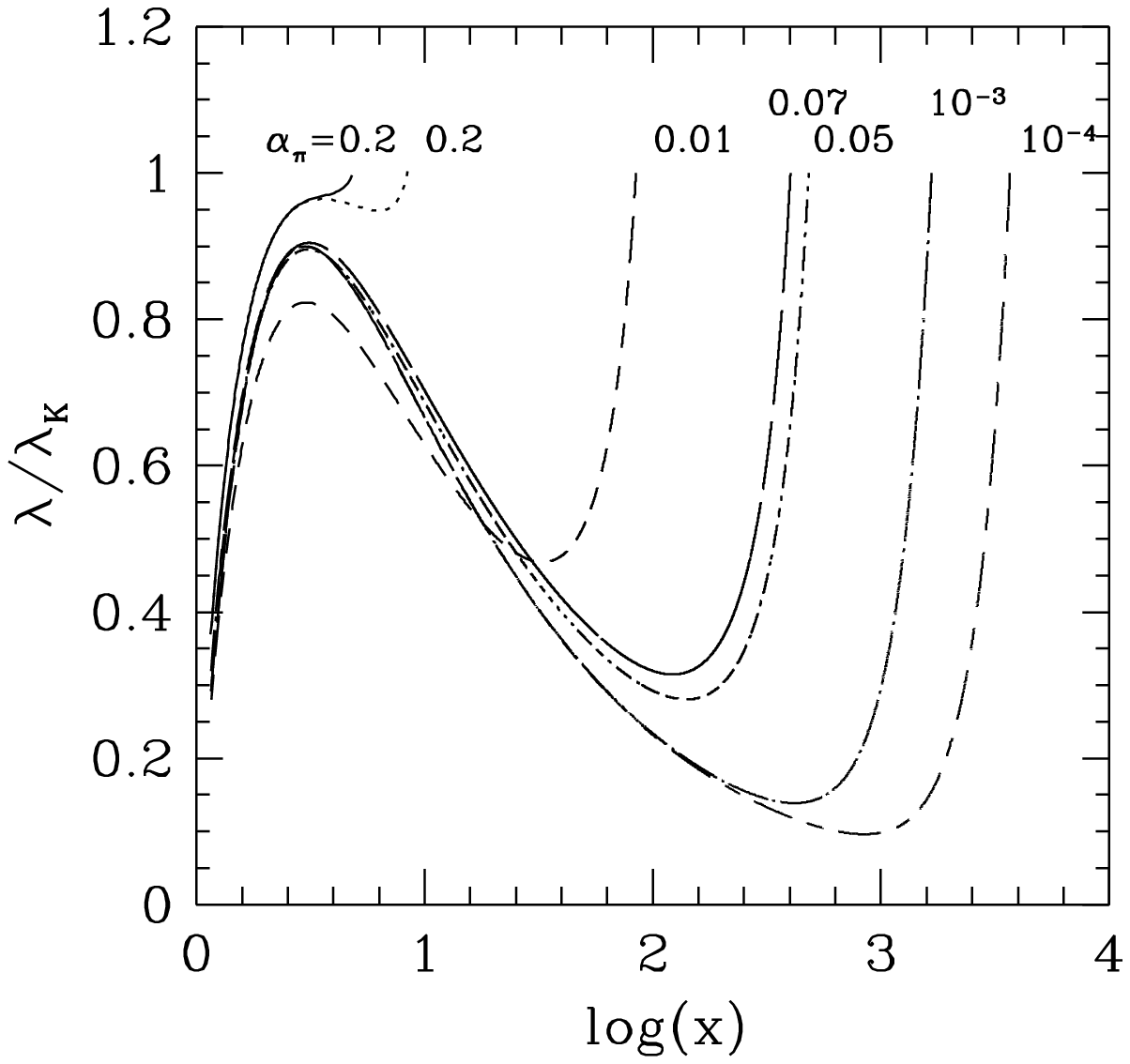,height=10truecm,width=10truecm,angle=0}}}
\vspace{0.5cm}
\noindent {\small {\bf Fig. 1.} Variation of $\lambda/\lambda_K$ with logarithmic radial distance for
a few solutions which are employed to study nucleosynthesis. The viscosity parameter
$\alpha_\Pi$ is marked on each curve. $x=x_K$ where $\lambda/\lambda_K=1$, represents the
location where the flow deviates from a Keplerian disk.
Note that except for the dashed curve marked $0.01$ (which is for $\gamma=5/3$, and the rest are for
$\gamma=4/3$), $x_K$ generally rises with decreasing $\alpha_\Pi$. Thus,  high viscosity
flows must deviate from the Keplerian disk closer to the black hole. }
\end{figure}

In C90 and C96, it was pointed out that in a large region of the
parameter space, especially for intermediate viscosities, centrifugal-pressure-supported shocks would be present in the hot, accretion flows.
In these cases a shock-free solution passing through the outer
sonic point was present. However, this branch is not selected by the flow and the flow
passes through the higher entropy solution through shocks and the
inner sonic points instead. This assertion has been repeatedly verified 
independently by both theoretical (Yang \& Kafatos 1995, Nobuta \& 
Hanawa 1994; Lu \& Yuan 1997; Lu et al. 1997) and numerical 
simulations (with independent codes, Chakrabarti \& Molteni 1993; 
Sponholz \& Molteni 1994; Ryu et al. 1995, Molteni et al.
1996 and references therein). When the shock forms, the temperature of 
the flow suddenly rises and the flow slows down considerably, 
raising the residence time of matter significantly. This effect of 
shock-induced nucleosynthesis is also studied in the next section 
and, for comparison, the changes in composition in the
shock-free branch were also computed, although it is understood that
the shock-free branch is unstable. Our emphasis is not on
shocks per se, but on the centrifugal-pressure-dominated region where the
accreting matter slows down. When the shock does not form,
the rise in temperature is more gradual. We generally 
follow the results of CT95 and C97 to compute  the temperature 
of the Comptonized flow in the sub-Keplerian region which may or may 
not have shocks. Basically we borrow the mean factor 
$F_{\mathrm{Compt}} \lsim 1$ by which the temperature of 
the flow at a given radius $x$ ($<x_{K}$) is reduced 
due to Comptonization process from the value dictated by the 
single-temperature hydrodynamic equations. This factor is typically $1/30 \sim 0.03$ 
for very low ($\lsim 0.1$) mass accretion rate of the {\it Keplerian 
component} (which supplies the soft photons for the Comptonization) 
and around $1/100 \sim 0.01$ or less for higher Keplerian
accretion rates. In presence of magnetic fields, some dissipation is
present due to reconnections. Its expression is $Q_{\mathrm{mag}}=\frac{3 B^2}{16\pi x \rho} v$
(Shvartsman 1971; Shapiro 1973). We do not assume this heating in this paper.

The list of major nuclear reactions such as PP 
chain, CNO cycle, rapid proton capture and alpha ($\alpha$) 
processes, photo-dissociation etc. which may take place inside a disk 
are already given in CJA, and we do not repeat them here. Suffice 
it to say that due to the hotter nature of the sub-Keplerian disks,
especially when the accretion rate is low and Compton cooling 
is negligible, the major process of hydrogen burning is some rapid proton 
capture process (which operates at $T \gsim 0.5 \times 10^9$K) and mostly ($p,\alpha$) reactions
as opposed to the PP chain (which operates at much lower temperature $T \sim 0.01-0.2 
\times 10^9$K) and CNO cycle (which operates at $T \sim 0.02-0.5 \times 10^9$K) as in CJA. 

Typically, accretion onto a stellar-mass black hole takes place 
from a binary companion which could be a main sequence star. In a
supermassive black hole at a galactic center, matter is presumably 
supplied by a number of nearby stars. Because it is difficult 
to establish the initial composition of the inflow, we generally take the solar
abundance as the abundance of the Keplerian disk. Furthermore, 
the Keplerian disk being cooler, and the residence time inside it being 
insignificant compared to the hydrogen burning time scale, we assume that
for $x \gsim x_{K}$, the composition of the gas remains the same as that of the 
companion star, namely, solar. Thus our computation starts only from the
time when matter is launched from the Keplerian disk. Occasionally, for comparison,
we run the models with an initial abundance same as the output of
big-bang nucleosynthesis (hereafter referred to as `big-bang abundance'). 
These cases are particularly relevant for nucleosynthesis
around proto-galactic cores and the early phase of star formations. We have also 
tested our code with an initial abundance same as the composition of late-type stars
since in certain cases they are believed to be companions of galactic 
black hole candidates (Martin et al. 1992, 1994; Filippenko et al.
1995; Harlaftis et al. 1996). 

\subsection {Selection of Models}

In selecting models for which the nucleosynthesis should be studied, the
following considerations were made. According to CT95, and C97, there
are two essential components of a disk. One is Keplerian (of rate ${\dot m}_d$)
and the other is sub-Keplerian halo (of rate ${\dot m}_h$).
For  ${\dot m}_d \lsim 0.1$ and ${\dot m}_h \lsim 1$, the black 
hole remains in hard states. A lower Keplerian accretion rate {\it generally} 
implies a lower viscosity and a larger $x_K$ ($x_K \sim 30-1000$; see, 
C96 and C97). In this parameter range the protons remain hot, 
typically, $T_p \sim 1-10 \times 10^9$ degrees or so. This is because the
efficiency of emission is lower ($f=1-Q^-/Q^+\sim 0.1$, where,
$Q^+$ and $Q^-$ are the height-integrated heat generation and heat loss 
rates [ergs cm$^{-2}$ sec$^{-1}$] respectively.
Also, see Rees (1984), where it is argued that ${\dot m}/\alpha^2$ is a
good indication of the cooling efficiency of the hot flow.). Thus, we 
study a group of cases (Group A) where the net accretion rate ${\dot m}
\sim 1.0$ and the viscosity parameter $\alpha \sim 0.001-0.1$.
The Comptonization factor $F_{\mathrm{Compt}} \sim 0.03$, i.e., the cooling due
to Comptonization reduces the mean temperature roughly by a factor of
around $30$, which is quite reasonable. Here, although the density of 
the gas is low, the temperature is high enough to cause significant nuclear 
reactions in the disk.

When the net accretion rate is very low (${\dot m} \lsim 0.01$) such 
as in a quiescence state of an X-ray novae, the dearth of soft photons 
keeps the temperature of the sub-Keplerian flow to a very 
high value and a high Comptonization factor $F_{\mathrm{Compt}}\sim 0.1$ could 
be used (Group B). Here significant nuclear reaction takes place, even 
though the density of matter is very low. Basically, the entire amount of
matter is photo-dissociated into protons and neutrons in this case
even when  opacity is very low.

In the event the inflow consist of both the Keplerian (accretion rate 
$\dot m_d$) and sub-Keplerian (accretion rate ${\dot m}_h$)
matter as the modern theory predicts, there would be situations where 
the {\it net} accretion rate is high, say ${\dot m}= {\dot m}_d 
+ {\dot m}_h \sim 1-5$, and yet the gas temperature is very high 
($T > 10^9$). This happens when viscosity is low
to convert sub-Keplerian inflow into a Keplerian disk.
Here, most of the inflow is in the sub-Keplerian component and very 
little (${\dot m}_d \sim 0.01$) matter is in the Keplerian flow.
Dearth of soft photons keeps the disk hot, while the density 
of reactants is still high enough to have profuse nuclear reactions. The simple criteria
for the cooling efficiency (that ${\dot m}/\alpha^2 >1$ would cool the disk, see Rees 1984)
will not hold since the radiation source (Keplerian disk) is different from the cooling body
(sub-Keplerian disk).

One could envisage yet another set of cases (Group C), where the accretion rate
is very high (${\dot m} \sim 10-100$), and the soft photons are so profuse
that the sub-Keplerian region of the disks becomes very 
cold. In this case, typically, viscosity is very high $0.2$, 
$x_K$ becomes low ($x_K \sim 3-10$). The efficiency 
of cooling is very high ($Q^+ \approx Q^-$, i.e., $f\approx 0$). 
The Comptonization factor is low $F_{\mathrm{Compt}} \lsim 0.01$. The black hole is 
in a soft state. There is no significant nuclear reaction in these cases.
In the proto-galactic phase when the supply of matter is 
{\it very} high, while the viscosity may be so low (say, $10^{-4}$) that 
the entire amount is not accreted, one can have an ultra-cold accretion flow 
with $F_{\mathrm{Compt}} \sim 10^{-3}$. In this case also not much nuclear reaction goes on.

The above simulations have been carried out with polytropic index $\gamma=4/3$. 
In reality, the polytropic index could be in between $4/3$ and $5/3$. 
If $\gamma <1.5$ then shocks would form as in  some of the above cases. 
However, for $\gamma >1.5 $, standing shocks would not form (C96). We have included 
one illustrative example of a shock-free case with $\gamma=5/3$ which is very
hot and we have presented the result in Group B. In this case 
the Keplerian component is far away and the intercepted soft photons are very few.

\subsection{Selection of the Reaction Network}

In selecting the reaction network we kept in mind the fact that hotter
flows may produce heavier elements through triple-$\alpha$ and 
proton and $\alpha$ capture processes. Similarly, due to photo-dissociation,
significant neutrons may be produced. Thus, we consider a sufficient number
of isotopes on either side of the stability line. The network thus
contains protons, neutrons, till $^{72}Ge$ -- altogether 255 nuclear 
species. The network of coupled non-linear differential equation is
linearized and evolved in time along the solution of C96 obtained from
a given set of initial parameters of the flow. This well proven method 
is widely used in the literature (see Arnett \& Truran 1969; Woosley et al.  1973).

The reaction rates were taken from Fowler et al. (1975)
including updates by Harris et al. (1983). Other relevant references
from where rates have been updated are: Thielemann (1980); Wallace
\& Woosley (1981); Wagoner et al.(1967); Fuller et al.(1980, 1982). For details of the procedure of adopting
reaction rates, see, CJA and Jin et al.(1989, hereinafter JAC). 
The solar abundance which was
used as the initial composition of the inflow was taken from Anders
\& Ebihara (1982). 

\section{Results}

In this section, we present a few major results of our simulations using 
different parameter groups as described above. For a complete solution of the 
sub-Keplerian disks (C96) we need to provide 
(a) the mass of the black hole $M$, (b) the viscosity
parameter $\alpha_\Pi$, (c) the cooling efficiency factor $f$, (d) the
Comptonization factor $F_{\mathrm{Compt}}$, (d) the net accretion rate of the
flow ${\dot m}$, (e) the inner sonic point location $x_{in}$ through which 
the flow must pass and finally, (f) the specific angular momentum 
$\lambda_{\mathrm{in}}$ at the inner sonic point. 

The following table gives the cases we discuss in this paper. 
The $\Pi$-stress viscosity parameter $\alpha_\Pi$, the location of the inner sonic 
point $x_{\mathrm{in}}$ and the value of the specific angular
momentum at that point $\lambda_{\mathrm{in}}$ are free parameters.
The net accretion rate ${\dot m}$, the Comptonization factor $F_{\mathrm{Compt}}$
and the cooling efficiency $f$ are related  quantities (CT96, C97).
For extremely inefficient cooling, $f\sim 1.0$, and for extremely efficient
cooling $f=0$ or even negative. The derived quantities, such as the value of maximum 
temperature $T_9^{\mathrm{max}}$ of the flow (in units of $10^9$K), density of matter (in cgs units) at $T_9^{\mathrm{max}}$, 
$x_K$, the location from where the Keplerian disk on the
equatorial plane becomes sub-Keplerian are also provided in the table. In the rightmost
column, we present whether the inner edge of the disk is stable (S) or unstable (U) in the presence of
the accretion flow. Three groups are separated as the parameters are clearly from three distinct regimes.


\bigskip

\centerline{ TABLE 1}

\begin{center}
\begin{tabular}{lllllllllllll}
\hline
\hline
Model & $M/M_\odot$ & $\gamma$ &$x_{\mathrm{in}}$ &$\lambda_{\mathrm{in}}$ & $\alpha_\Pi$& ${\dot m}$&
f & $F_{\mathrm{Compt}}$ & $x_K$ &$T_9^{\mathrm{max}}$&$\rho_{\mathrm{max}}$&S/U\\
\hline
\hline
A.1&10& 4/3 & 2.7945 & 1.65& 0.001& 1& 0.1& 0.03& 1655.7 &5.7&6.2$\times 10^{-7}$&S\\
A.2&10& 4/3& 2.9115&1.6&0.07&1& 0.1&0.03&401.0 &4.7& 4.9$\times 10^{-7}$&S\\
A.3&$10^6$& 4/3&2.9115& 1.6&0.07& 1 &0.1&0.03&401.0 &4.7&4.9$\times 10^{-12}$&U\\
\hline
B.1 &10 & 4/3&2.8695&1.6&0.05&0.01&0.5&0.1& 481.4 &16.5& 3.9$ \times10^{-9}$&S\\
B.2 &10 & 4/3 &  2.8695& 1.6& 0.05&4&0.5&0.1&481.4 &16.5& 1.6$ \times 10^{-8}$&U\\
B.3 &10 & 5/3 & 2.4 &1.5 & 0.01 & 0.001  & 0.5 & 0.1 & 84.4 &47& 3.3$\times 10^{-10}$&S\\
B.4 &10 & 4/3 &  2.795& 1.65& 0.2&0.01  &0.2&0.1&8.4 &13&1.1$\times 10^{-8}$&S\\
\hline
C.1&10& 4/3 & 2.795& 1.65& 0.2& 100 &0.0&0.01&4.8 &0.8&1.1$ \times 10^{-4}$&S\\
C.2&$10^6$ & 4/3& 2.795 &1.65& $10^{-4}$ &$100$ &0.0&0.001&3657.9 &0.2& 6.2$\times 10^{-10}$&S\\
\hline
\hline
\end{tabular}
\end{center}

The basis of our three groupings are clear from the Table. Very 
low ${\dot m}/\alpha_\Pi^2$ in Group B makes the cooling efficiency 
to be very small. Thus we choose a relatively large $f \sim 0.2-0.5$. 
It also makes the cooling due to Comptonization to be very  low
($F_{\mathrm{Compt}}  \sim 0.1$). Thus the disks could be ultra-hot. Intermediate ${\dot m}/\alpha_\Pi^2$
in Group A means that the efficiency of cooling is intermediate $f \sim 0.1$ and the 
Compton cooling of the sub-Keplerian region is average: $F_{\mathrm{Compt}} \sim 0.03$. 
The sub-Keplerian disk in this case is neither 
too hot nor too cold. Extremely high ${\dot m}/\alpha_\Pi^2$ 
causes a strong cooling in Group C. Thus, we choose $f=0$, and a very efficient 
Compton cooling $F_{\mathrm{Compt}}\sim 0.01-0.001$. As a result,
the disk is also very cold. Now, we present our numerical results in these cases.

\subsection{Nucleosynthesis in Moderately Hot Flows}

\noindent {\it Case A.1}: In this case, the termination of the
Keplerian component in the weakly viscous flow takes place at
$x=1655.7$. The soft photons intercepted by the
sub-Keplerian region reduce the temperature of this
region but not by a large factor. The net accretion rate 
${\dot m}=1$ is the sum of (very low) Keplerian 
component and the sub-Keplerian component. Using computations of CT95
and C97 for ${\dot m}_d \sim 0.1$ and ${\dot m}_h \sim 0.9$, we find that 
the electron temperature $T_e$ is around $60$keV $T9\sim 0.6$ 
($T_9$ is the temperature in units of $10^9$K) and the ion temperature 
is around $T_9=2.5$. This fixes the Comptonization factor
to  about $F_{Compt} = 0.03$. This factor is used to reduce the
temperature distribution of solutions of C96 (which does not
explicitly use Comptonization) to temperature distribution {\it with} 
Comptonization. The ion temperature (in $T_9$) and density (in units of $10^{-10}$ gm cm$^{-3}$
to bring in the same plot) distribution computed in this manner 
are shown in Fig. 2a. Figure 2b gives the velocity distribution
(velocity is measured in units of $10^{10}$ cm sec$^{-1}$).
Note the sudden rise in temperature and slowing down of matter
close to the centrifugal barrier $x \sim 30$. Figure 2c shows the
changes in composition as matter is accreted onto the black hole.
Only those species with abundance $Y_i \gsim 10^{-4}$ have been
shown for clarity. Also, compositions closer to the 
black hole are shown, as variations farther out are 
negligible. Most of the burning of species takes place 
below $x=10$. A significant amount of the neutrons
(with a final abundance of $Y_n \sim 10^{-3}$) is produced 
by the photo-dissociation process. Note that closer to the black hole,
$^{12}\!C$, $^{16}\!O$, $^{24}\!Mg$ and $^{28}\!Si$ are all destroyed 
completely, even though at around $x=5$ or so, the abundance of some of them
went up first before going down. Among the new species which are formed 
closer to the black hole are $^{30}\!Si$, $^{46}\!Ti$, $^{50}\!Cr$. 
The final abundance of 
$^{20}\!Ne$ is significantly higher than the initial value. This was not dissociated
as the residence time in hotter region was insufficient. Thus a significant
metalicity could be supplied by winds from the centrifugal barrier.  

\begin {figure}
\vbox{
\vskip -5.0cm
\hskip 0.0cm
\centerline{
\psfig{figure=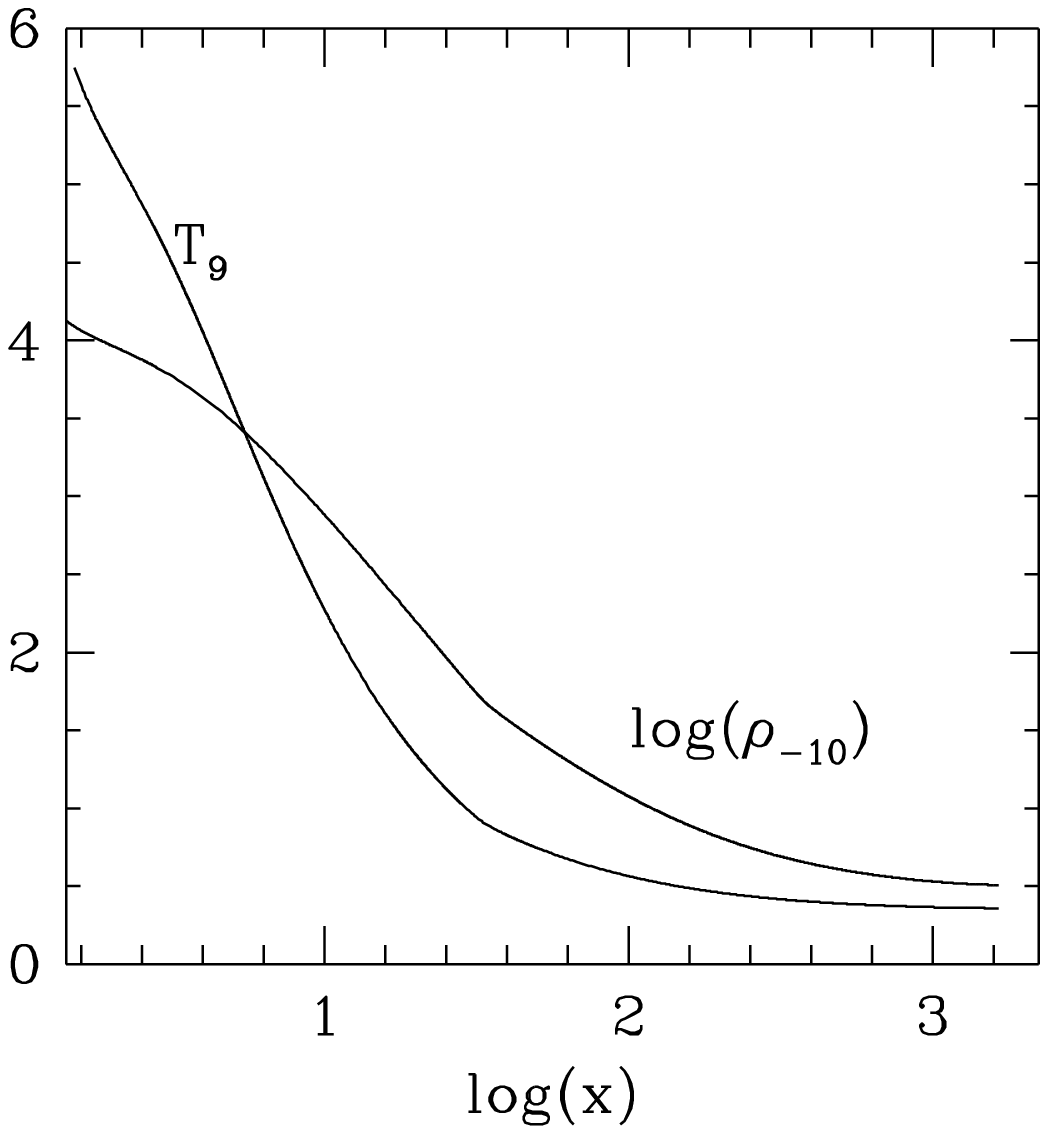,height=10truecm,width=10truecm,angle=0}}}
\vspace{-1.5cm}
\noindent{\small {\bf Fig. 2a}}
\end{figure}

\begin {figure}
\vbox{
\vskip -5.0cm
\hskip 0.0cm
\centerline{
\psfig{figure=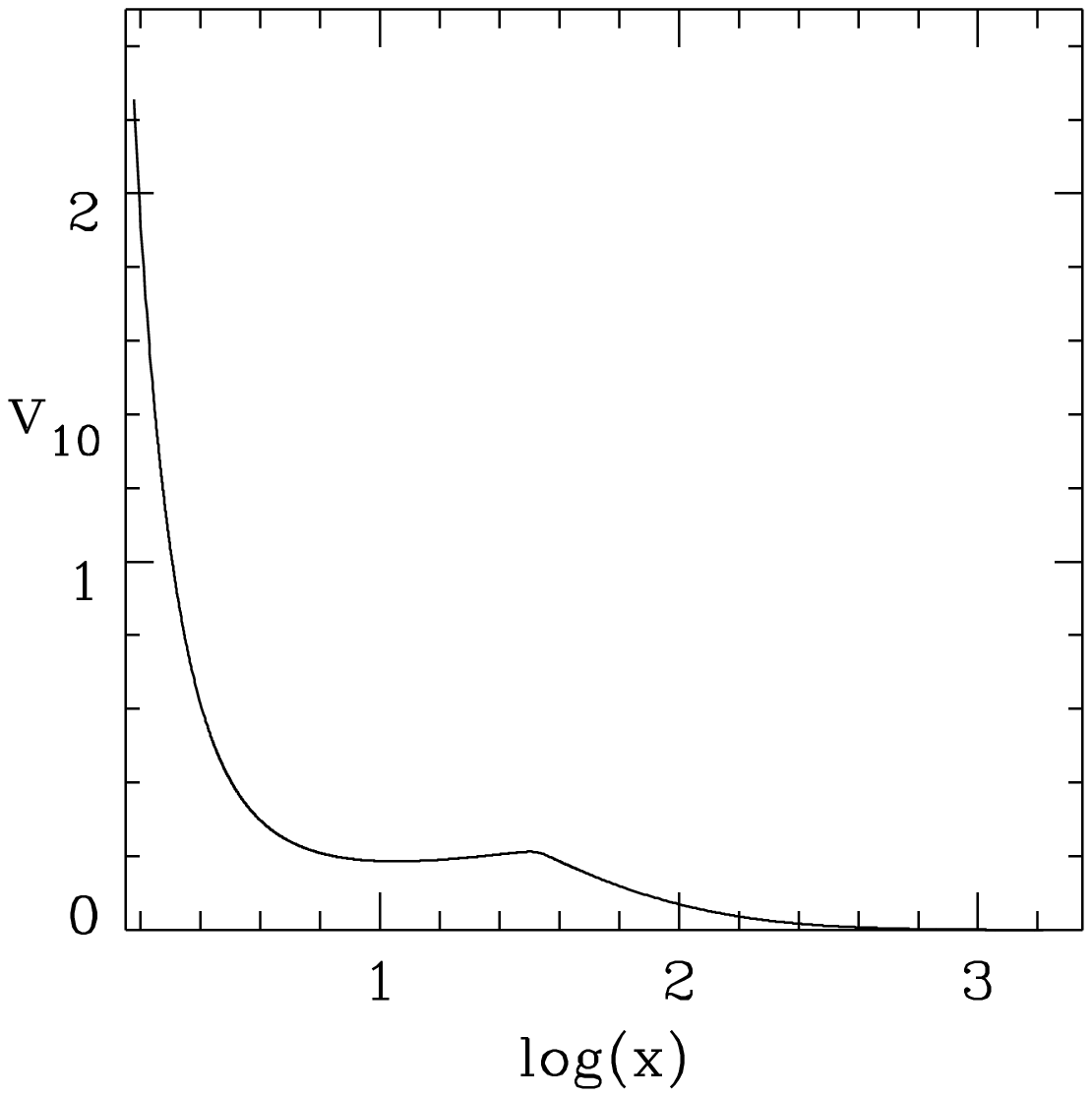,height=10truecm,width=10truecm,angle=0}}}
\vspace{-1.5cm}
\noindent{\small {\bf Fig. 2b}}
\end{figure}

\begin {figure}
\vbox{
\vskip -5.0cm
\hskip 0.0cm
\centerline{
\psfig{figure=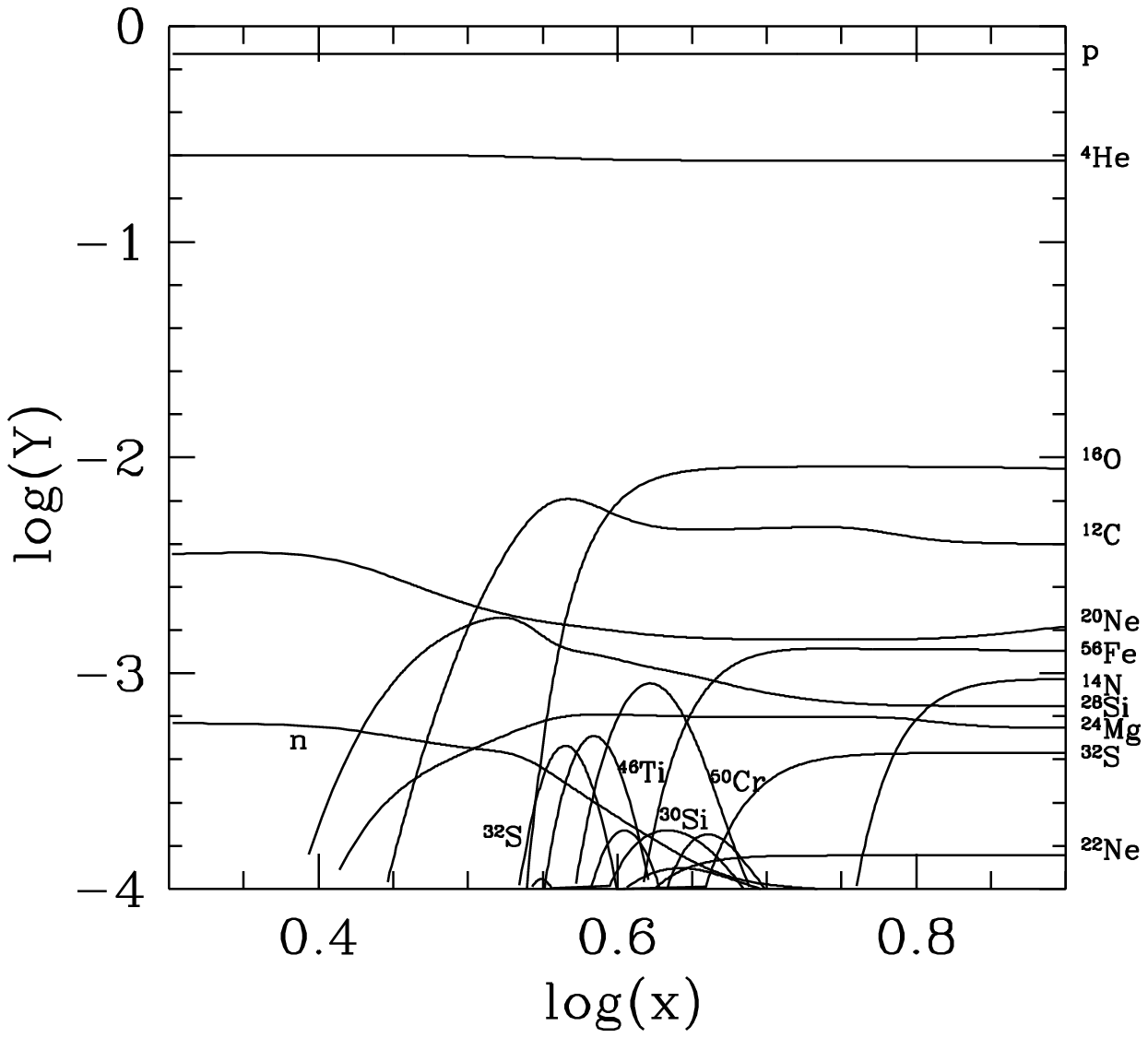,height=10truecm,width=10truecm,angle=0}}}
\vspace{-1.5cm}
\noindent{\small {\bf Fig. 2c}}

\end{figure}

\begin {figure}
\vbox{
\vskip -4.0cm
\hskip 0.0cm
\centerline{
\psfig{figure=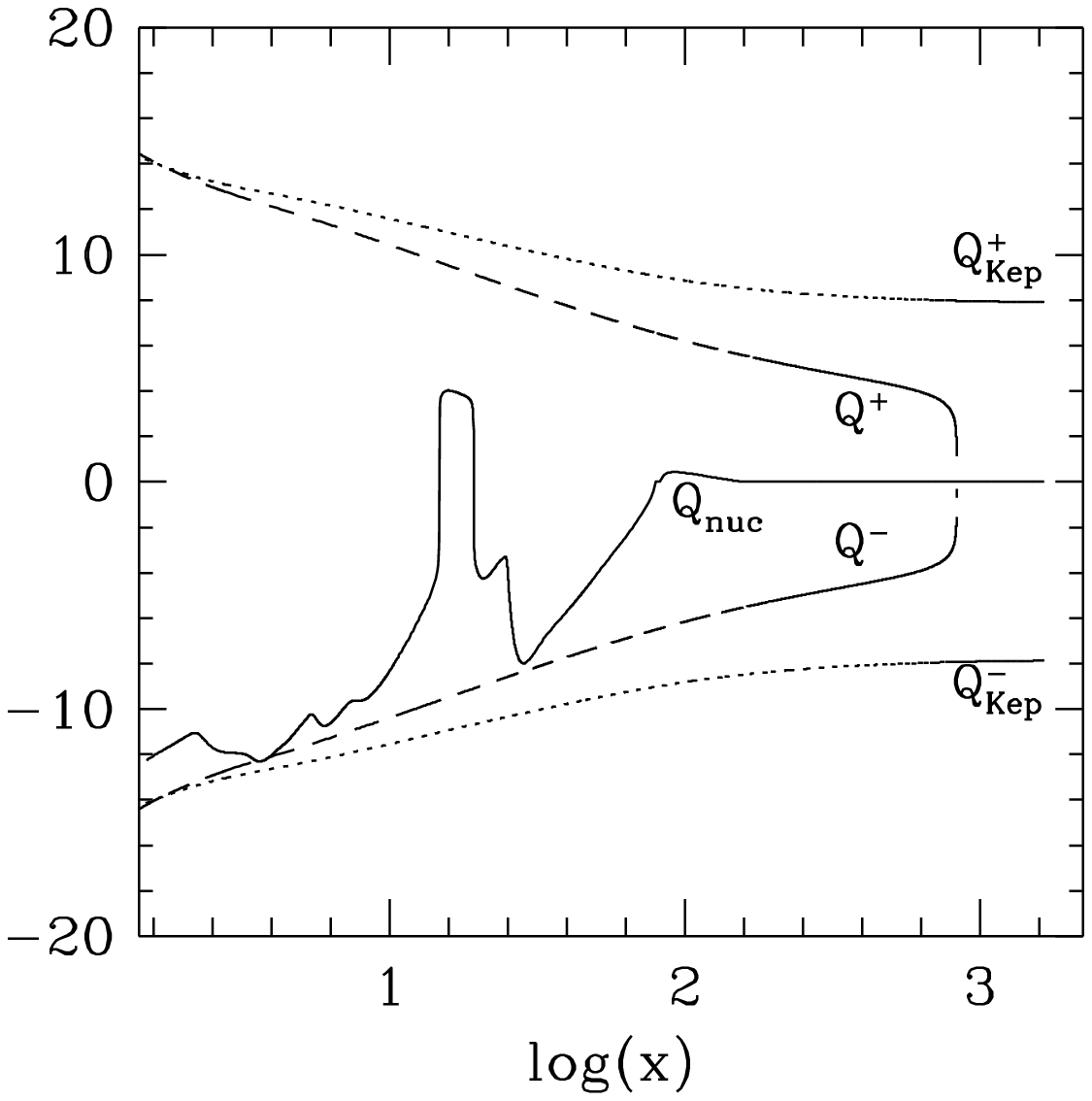,height=10truecm,width=10truecm,angle=0}}}
\vspace{-1.5cm}
\noindent{\small {\bf Fig. 2d}}

\end{figure}

\begin {figure}
\vbox{
\vskip 2.0cm
\noindent {\small {\bf Fig. 2.} Variation of (a) ion temperature ($T_9$) and
density ($\rho_{-10}$), (b) radial velocity $v_{10}$,
(c) matter abundance $Y_i$ in
logarithmic scale and (d) various forms of height-integrated specific energy release
and absorption rates (in ergs cm$^{-2}$ sec$^{-1}$) when the
model parameters are $M=10M_\odot$, ${\dot m}=1.0$, $\alpha_\Pi=0.001$
as functions of logarithmic radial distance ($x$ in units of Schwarzschild radius).
Q is in logarithmic scale. See text and Table 1 for other parameters of Case A.1 which 
is considered here. The centrifugal barrier slows down and heats up matter
where a significant change in abundance takes place ($\Delta Y_i \sim 10^{-3}$)}.
}
\end{figure}

Figure 2d shows the energy release and absorption due to exothermic and endothermic
nuclear reactions ($Q_{\mathrm{nuc}}$) that are taking place inside the disk (solid). Superposed on 
it are the energy generation rate $Q^+$ (long dashed curve) due to viscous process and the
energy loss rate $Q^-$ in the sub-Keplerian flows. For comparison, we also
plot the hypothetical energy generation and loss rates (short dashed curves marked as
$Q^+_{\mathrm{Kep}}$ and $Q^-_{\mathrm{Kep}}$ respectively) if the disk had purely Keplerian 
angular momentum distribution even in the sub-Keplerian regime.
All these quantities are in units of $3\times 10^6$ and they represent height-integrated
energy release rate (ergs cm$^{-2}$ sec$^{-1}$). Note that these Qs are in
logarithmic scale (if $Q<0$, $-log(|Q|)$ is plotted).
As matter leaves the Keplerian flow, the proton capture ($p, \alpha$) processes
(such as $^{18}\!O (p,\alpha) ^{15}\!N$, $^{15}\!N (p, \alpha) ^{12}\!C$,
$^{6}\!Li (p,\alpha) ^{3}\!He$, $^{7}\!Li (p,\alpha) ^{4}\!He $, 
$^{11}\!B(p,\gamma)3\alpha$, $^{17}\!O (p,\alpha)^{14}\!N$,  etc.)
burn hydrogen and release energy to the disk. (Since the
temperature of the disk is very high, PP chains or CNO cycles are not the dominant
processes for the energy release.)
At around $x=40$, the deuterium starts burning ($D(\gamma,n)p $)
and the endothermic reaction causes the nuclear energy release to become `negative', i.e.,
a huge amount of energy is absorbed from the disk. At the completion
of the deuterium burning (at around $x=20$) the energy release tends to goes back to the
positive value to the level dictated by the original proton capture processes. Excessive
temperature at around $x=5$ breaks $^3\!He$ down into deuterium 
($^3\!He(\gamma, p)D$, $D(\gamma,n)p$). Another major 
endothermic reaction  which is dominant in this region is
$^{17}\!O (\gamma,n)^{16}\!O$. These reactions absorb
a significant amount of energy from the flow. Note that the nuclear energy
release or absorption is of the same order as the energy release  due to 
viscous process. This energy was incorporated in computing 
thermodynamic quantities following these steps:

\noindent (a) Compute thermodynamic quantities without nuclear energy
\noindent (b) Run nucleosynthesis code and compute $Q_{\mathrm{nuc}}$\\
\noindent (c) Fit $Q_{\mathrm{nuc}}$ using piecewise analytical curves
and include this into the definition of $f$,\\
$$
f=1-\frac{Q^-}{Q^++Q_{\mathrm{nuc}}}
\eqno{(1)}
$$ 
\noindent (d) Do sonic point analysis once more using this 
extra heating/cooling term and compute thermodynamic quantities.\\
\noindent (e) Repeat from step (b) till the results converge.
In this case, there in virtually no difference in the solution
and the solution appear to be completely stable under nucleosynthesis.\\

\noindent{\it Case A.2:} Here we choose the same net accretion rate, but 
with a larger viscosity. As a result, the Keplerian component moves
closer. The Comptonization is still not very effective, and 
the flow is moderately hot as above with $F_{\mathrm{Compt}}=0.03$. The flow
deviates from a very hot (sufficient to cause the flow to pass through the
outer sonic point) Keplerian disk at $x_K=401.0$, and after 
passing through an outer sonic point at $x=50$, and through a shock
at $x_S=15$, the flow enters into the black hole through the
inner sonic point at $x=2.9115$. We show the results both for the shock-free branch 
(i.e., the one which passes through only the outer sonic point before plunging into the
black hole, dotted curves) and the shocked branch of the solution (solid curves). 
Figure 3a shows the comparison of the temperatures and densities
(scaled in the same way as in Fig. 2a). The temperature and density jump 
sharply at the shock. Figure 3b shows the comparison of the 
radial velocities. The velocity sharply drops at the shock.
Both of these effects hasten the nuclear burning in the case
which includes the shock. Figure 3c shows the comparison of the abundances of 
only those species whose abundances reached a value of at least $10^{-4}$. The 
difference between the shocked and the shock-free cases is that in the shock
case similar burning takes place farther away from the black hole
because of much higher temperature in the post-shock region.

\begin {figure}
\vbox{
\vskip -5.0cm
\hskip 0.0cm
\centerline{
\psfig{figure=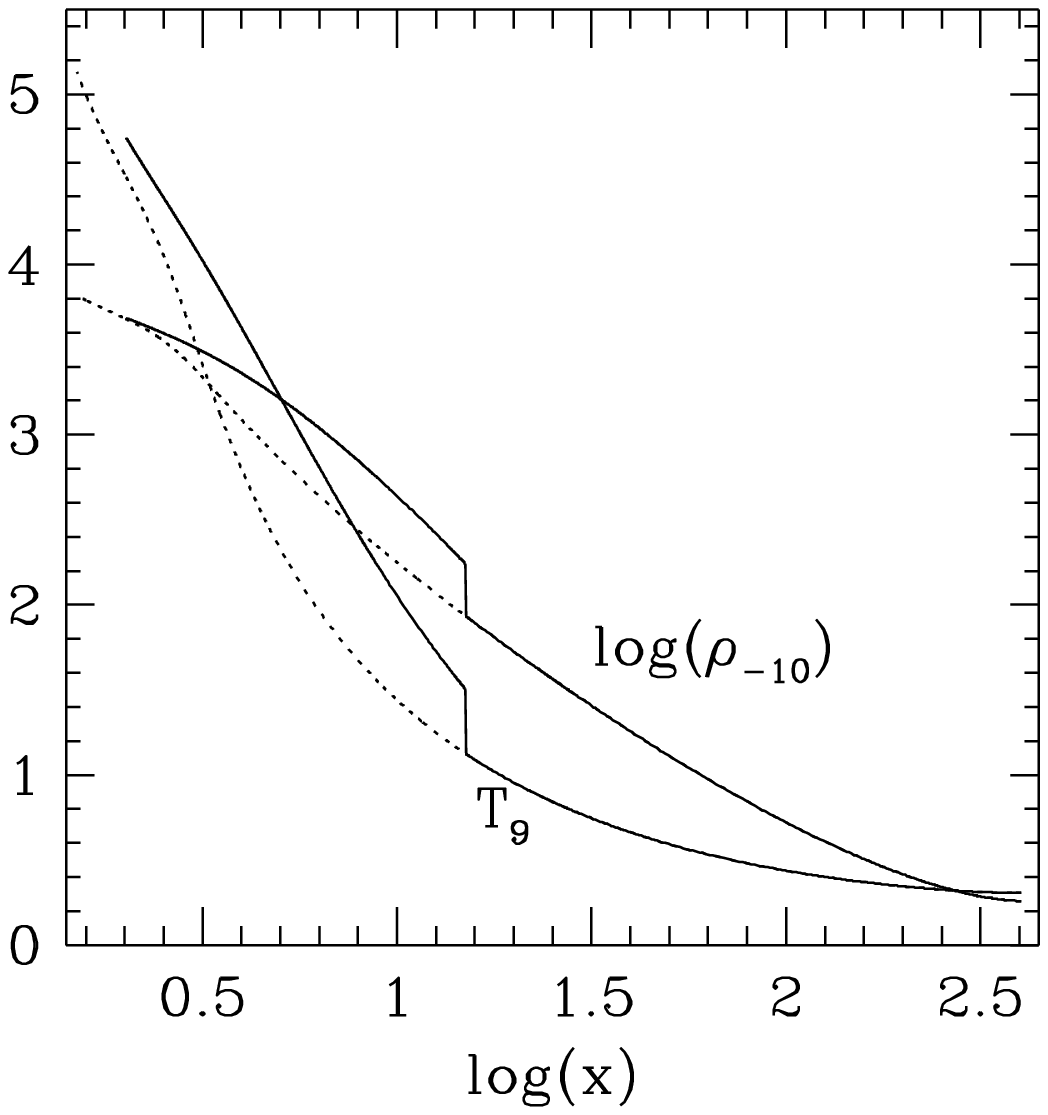,height=10truecm,width=10truecm,angle=0}}}
\vspace{-2.5cm}
\noindent{\small {\bf Fig. 3a}}

\end{figure}
\begin {figure}
\vbox{
\vskip -4.0cm
\hskip 0.0cm
\centerline{
\psfig{figure=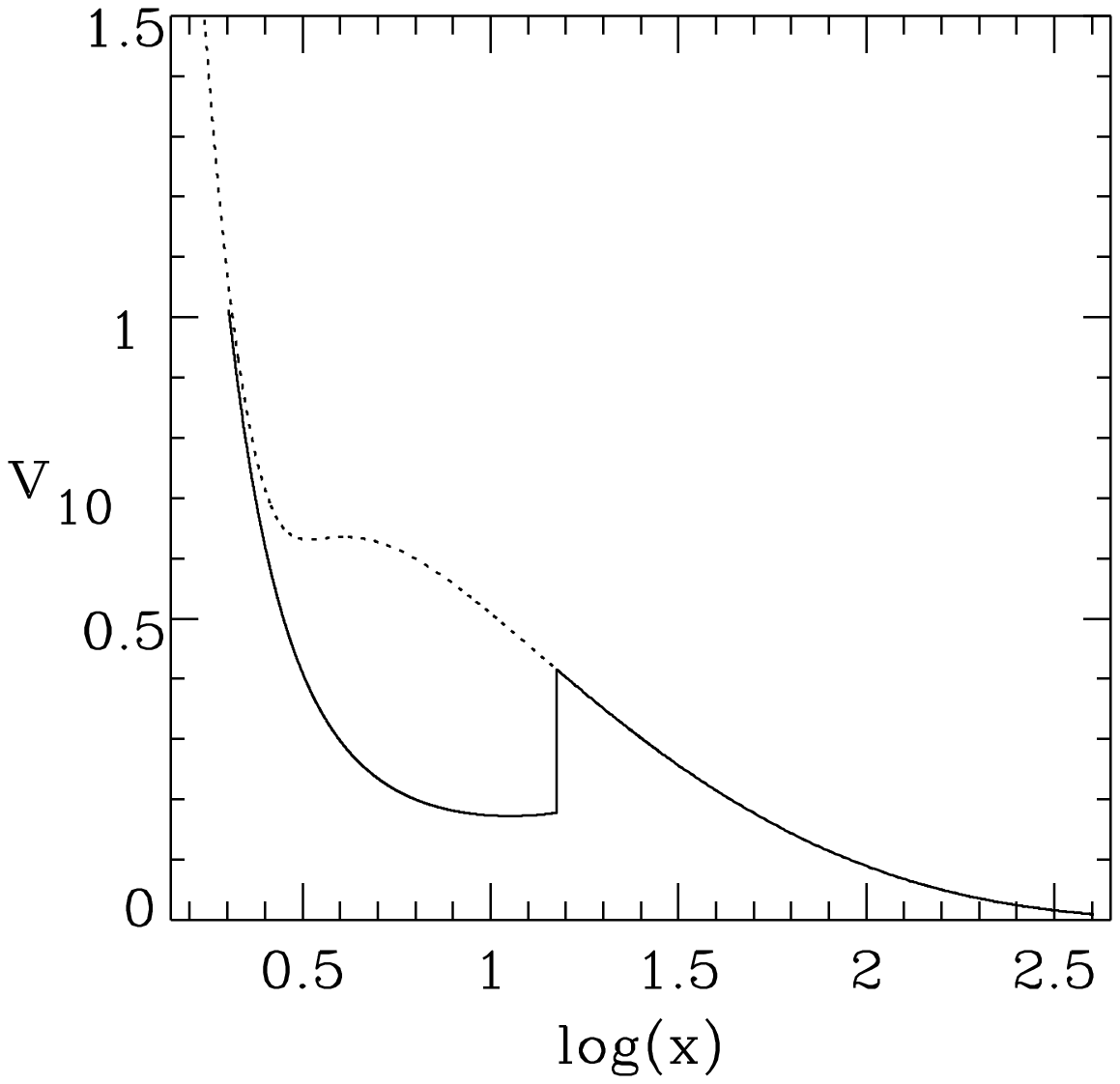,height=10truecm,width=10truecm,angle=0}}}
\vspace{-2.5cm}
\noindent{\small {\bf Fig. 3b}}
\end{figure}
\begin {figure}
\vbox{
\vskip -4.0cm
\hskip 0.0cm
\centerline{
\psfig{figure=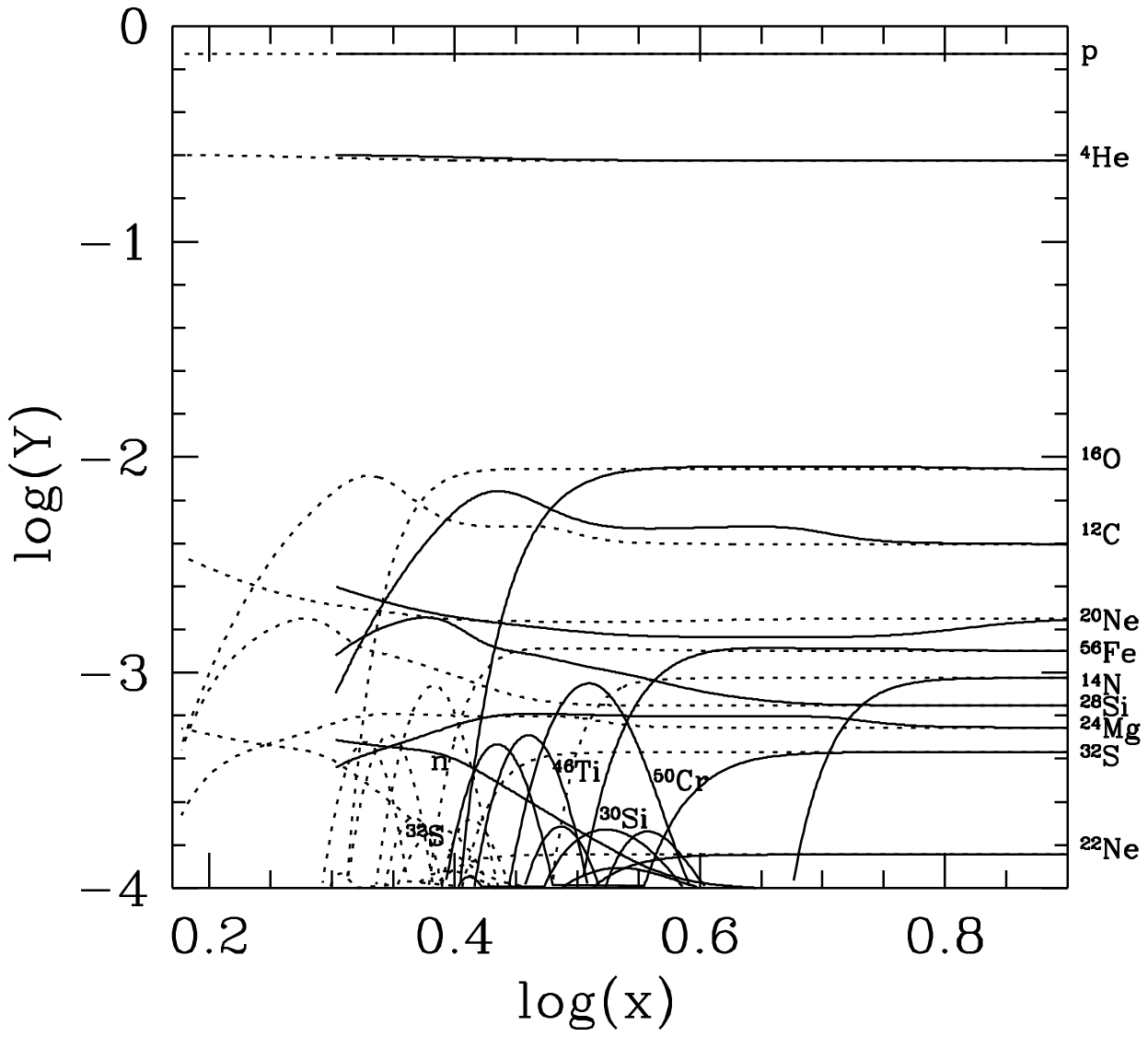,height=10truecm,width=10truecm,angle=0}}}
\vspace{-2.5cm}
\noindent{\small {\bf Fig. 3c}}

\end{figure}
\begin {figure}
\vbox{
\vskip -4.0cm
\hskip 0.0cm
\centerline{
\psfig{figure=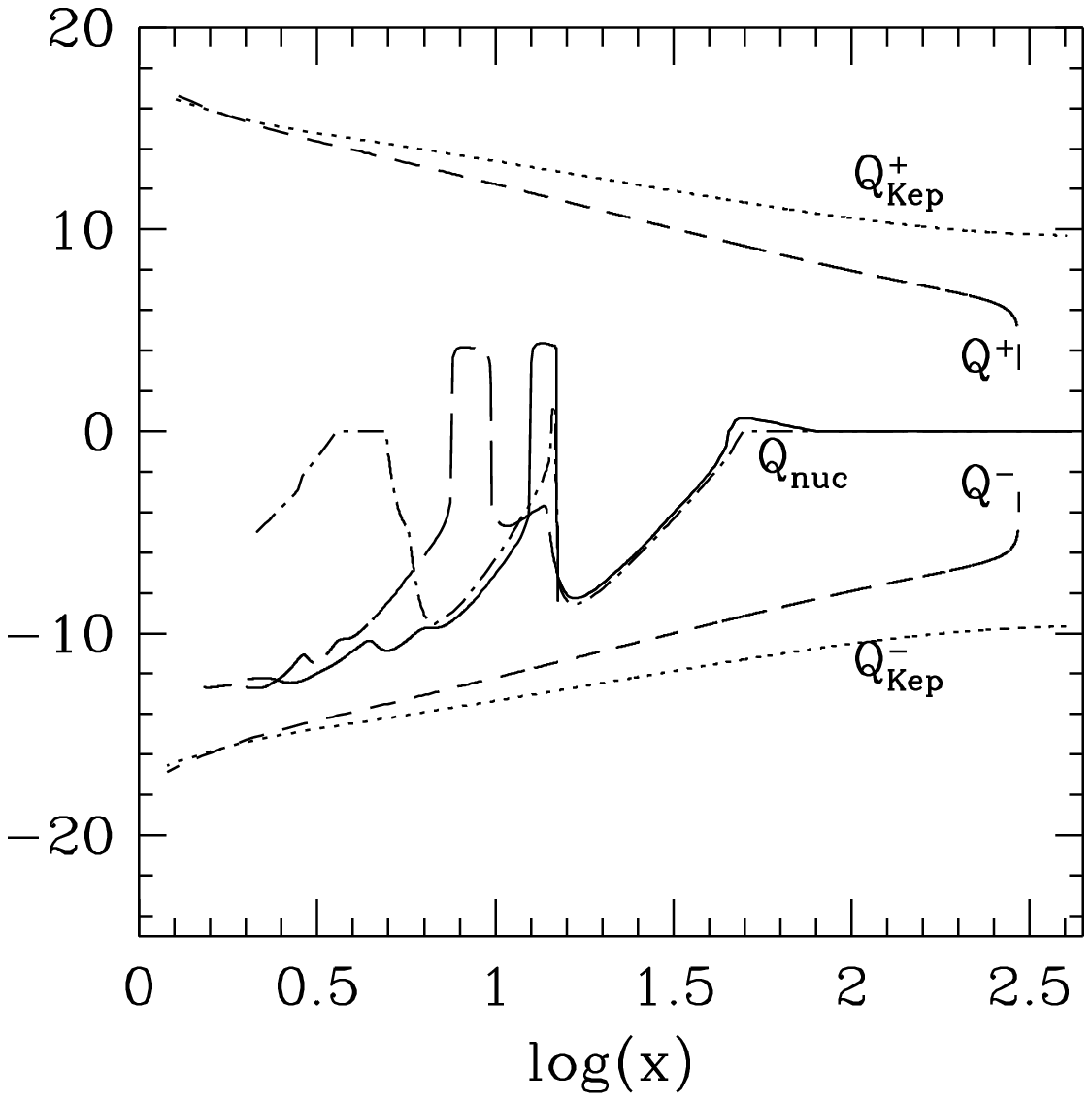,height=10truecm,width=10truecm,angle=0}}}
\vspace{-2.5cm}
\noindent{\small {\bf Fig. 3d}}
\end{figure}
\begin {figure}
\vbox{
\vskip 2.0cm
\noindent {\small {\bf Fig. 3.}: Variation of (a) ion temperature ($T_9$) and density ($\rho_{-10}$), (b) radial
velocity $v_{10}$, (c)  matter abundance $Y_i$ in logarithmic scale and 
(d) various forms of specific energy release
and absorption rates when the model parameters are $M=10M_\odot$, ${\dot m}=1.0$, $\alpha_\Pi=0.07$
as functions of logarithmic radial distance ($x$ in units of Schwarzschild radius).
See text and Table 1 for other parameters of Case A.2 which is considered here.
Solutions in the stable branch with shocks are solid curves and those without the shock
are dotted in (a-c). Curves in (d) are described in the text.
At the shock temperature and density rise significantly and cause
a significant change in abundance even farther out. Shock-induced winds may
cause substantial contamination of the galactic composition when parameters
are chosen from these regions.}}
\end{figure}

The nature of the (height integrated) nuclear energy release is very similar to Case A.1
as the major reactions which take place inside the disk are basically same,
except that the exact locations where any particular reactions take place are
different since they are temperature sensitive. In Fig. 3d, we
show all the energy release/absorption components for the shocked flow (solid curve).
For comparison, we include the nuclear energy curve of the shock-free branch 
(very long dashed curve). Note that in the post-shock region, hotter and denser flow 
of the shocked-branch causes a particular nuclear reaction to take place farther 
away from a black hole when compared with the behaviour in the shock-free branch 
as is also reflected in the composition variation in Fig. 3c. The viscous energy 
generation ($Q^+$) and the loss of energy ($Q^-$) from the disk (long dashed)
and shown. As before, these quantities, if the inner part had Keplerian
distribution, are also plotted (short dashed). When big-bang 
abundance is chosen to be the initial abundance, the net composition does 
not change very much, but the dominating reactions themselves are 
somewhat different because the initial compositions are different.
The dot-dashed curve shows the energy release/absorption in the shocked flow when 
big-bang abundance is chosen. All these quantities are, as before, in units of $3\times 10^6$ 
and they represent height integrated energy release rate (ergs cm$^{-2}$ sec$^{-1}$).
For instance, in place of proton capture
reactions for computations with solar abundance, the fusion of deuterium 
into $^4\!He$ plays a dominant role via the following reactions: 
$D(D,n)^3\!He$, $D(p,\gamma)^3\!He$, $D(D,p)T$, $^3\!He(D,p)^4\!He$. 
This is because no heavy elements were present 
to begin with and  proton capture processes involving heavy elements such as were prevalent 
in the solar abundance case cannot take place here. Endothermic reactions
at around $x=20-40$ are dominated by deuterium dissociation as before. However,
after the complete destruction of deuterium, the exothermic reaction is momentarily
dominated by neutron capture processes (due to the same neutrons which are produced 
earlier via $D(\gamma,n)p$) such as $^3\!He(n,p)T$ which produces the 
spike at around $x=14.5$. Following this, $^3 \!He$ and $T$ are destroyed as in the  solar abundance case
(i.e., $^3\!He(\gamma,p)D$, $D(\gamma,n)p$, $T(\gamma,n)D$) and reaches the minimum
in the energy release curve at around $x=6$. The tendency of going back to the
exothermic region is stopped due to the photo-dissociation of $^4\!He$ via
$^4\!He(\gamma,p)T$ and $^4\!He(\gamma,n)^3\!He$. At the end of the 
big-bang abundance calculation, a significant amount of neutrons are produced.
The disk was found to be perfectly stable under nuclear reactions.

\noindent {\it Case A.3:} This case is exactly same as A.2 except that 
the mass of the black hole is chosen to be $10^6M_\odot$. 
The temperature and velocity variations are similar to the above 
case. Because the accretion rate (in non-dimensional units) is the same,
the density (which goes as ${\dot m}/r^2 v$) is lower by a factor of $10^{-5}$. 
Tenuous plasma should change its composition significantly only at higher 
temperatures than in the previous case. However, the increase in residence time 
by a factor of around $10^5$ causes the nuclear burning to take place 
farther out even at a lower temperature. This is exactly what is seen.
Figure 4a shows the comparison (without including nuclear energy) 
of the composition of matter when the
flow has a shock (solid curves) and when the flow is shock-free 
(dashed curve). We recall that the shock-free  flow is in reality not 
stable. It is kept only for comparison purposes. Note that unlike earlier cases,
a longer residence time also causes to burn all the $^{20}\!Ne$ that was generated
from $^{16}\!O$.

\begin {figure}
\vbox{
\vskip -4.5cm
\hskip 0.0cm
\centerline{
\psfig{figure=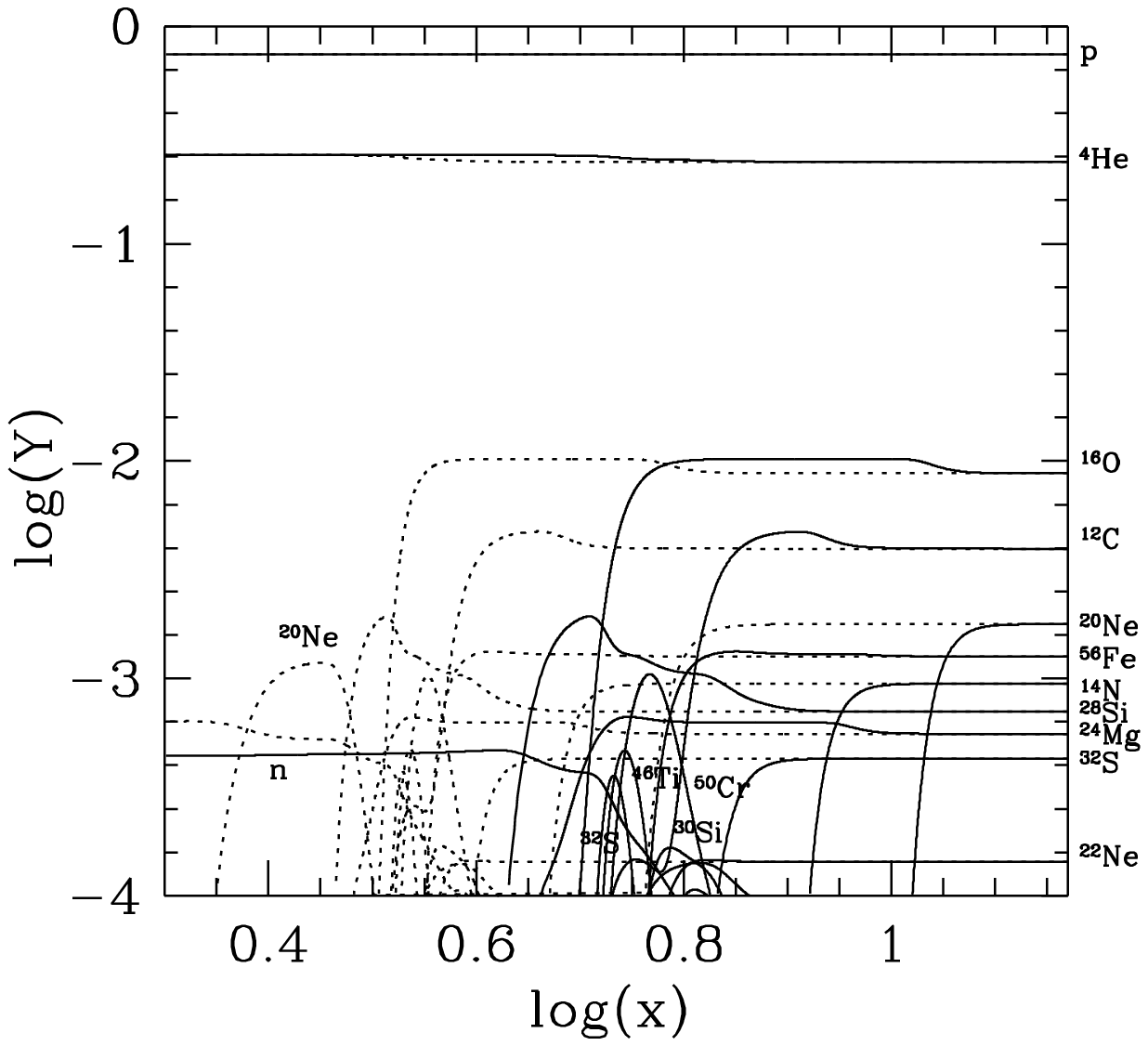,height=10truecm,width=10truecm,angle=0}}}
\vspace{-0.5cm}
\noindent{\small {\bf Fig. 4a}}
\end{figure}
\begin {figure}
\vbox{
\vskip -4.5cm
\hskip 0.0cm
\centerline{
\psfig{figure=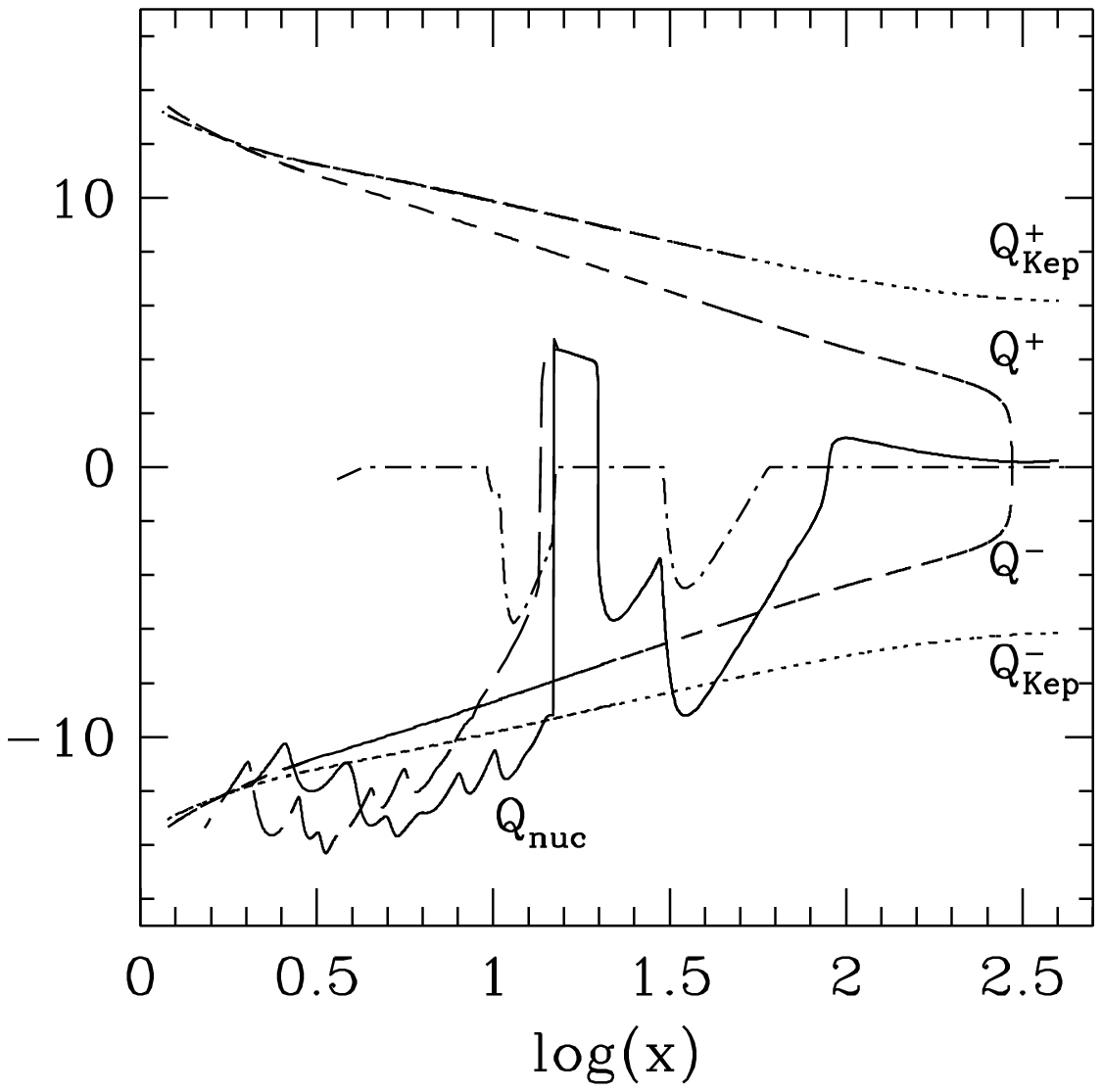,height=10truecm,width=10truecm,angle=0}}}
\vspace{-0.5cm}
\noindent{\small {\bf Fig. 4b}}
\end{figure}
\begin {figure}
\vbox{
\vskip -4.5cm
\hskip 0.0cm
\centerline{
\psfig{figure=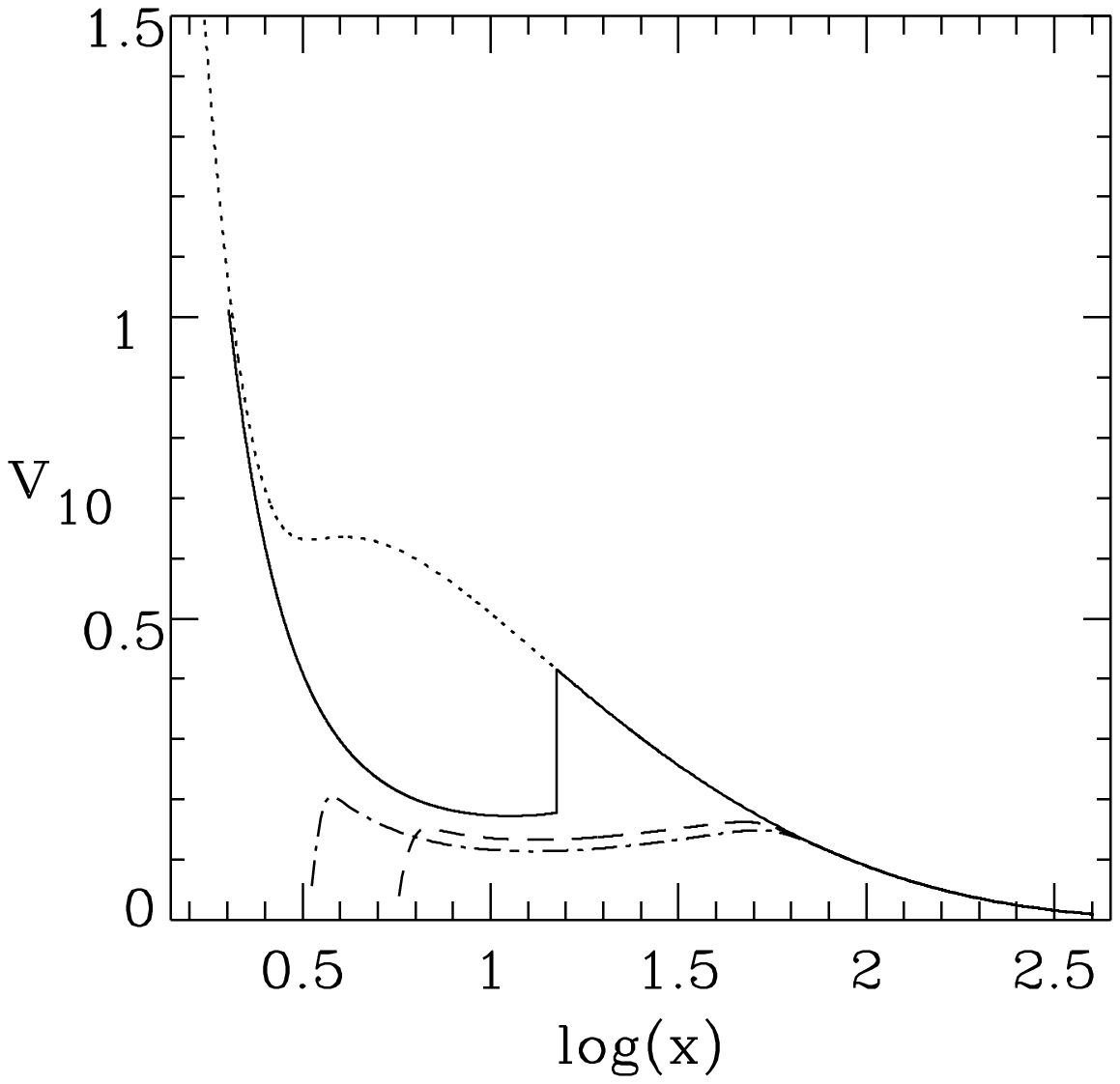,height=10truecm,width=10truecm,angle=0}}}
\vspace{-0.5cm}
\noindent{\small {\bf Fig. 4c}}
\end{figure}

\begin {figure}
\vbox{
\vskip 2.0cm
\noindent {\small {\bf Fig. 4.}: Variation of (a)  matter abundance  $Y_i$ in logarithmic scale,
(b) various forms of height-integrated specific energy release and absorption rates and (c) velocity
(in units of $10^{10}$cm s$^{-1}$) when the model parameters
are $M=10^6 M_\odot$, ${\dot m}=1.0$, $\alpha_\Pi=0.07$
as functions of logarithmic radial distance ($x$ in units of Schwarzschild radius).
See text and Table 1 for other parameters of Case A.3.
In (a) solutions in the stable branch with shocks are solid curves and those without the shock
are short dashed. Curves in (b) are described in the text.
Basic conclusions are as in the previous case. In (c), dot-dashed curve and dashed curves
are drawn when nuclear energy is taken into account.}}

\end{figure}

In Fig. 4b, we show a comparison of various height-integrated energy release and absorption curves 
as in Fig. 3d (in ergs cm$^{-2}$ sec$^{-1}$). The nuclear energy remains negligibly small till around $x=100$.
After that the endothermic reactions dominate. This is due to the dissociation of 
$D$, $^3\!He$ and $^7\!Li$ and also of $^{12}\!C$, $^{16}\!O$, $^{20}\!Ne$ etc. 
all of which produce $^4\!He$.  The solid curve is for the branch with a shock and the
very long dashed curve is for the shock-free branch. A small amount of neutrons are produced 
($Y_n \sim 10^{-3}$) primarily due to the dissociation of $D$. These
considerations are valid for solar abundance as the initial composition.
In the case of big-bang abundance (dash-dotted curve), similar reactions take place but 
no elements heavier than $^7\!Li$ are involved. The three successive dips
are due to dissociation of $D$, $^3\!He$ and $^4\!He$ respectively. 

Below $x=10$, $|Q_{\mathrm{nuc}}|$ is larger compared to $Q^+$
by 3-4 orders of magnitude. This is because of the superposition of 
a large number of photo-dissociation effects. We expect that 
in this case the disk would be unstable. This is exactly what we see.
In Fig. 4c, we show the effects of nuclear reactions more clearly. The dotted
curve and the solid curves are, as in Fig. 3b, the variation of velocity 
for the solution without and with shocks, respectively. The dot-dashed curve 
represents velocity variation without shock when nuclear reaction is
included. The dashed curve is the corresponding solution when nucleosynthesis of the
shocked branch is included.  Both branches are unstable
since the steady flow is subsonic at the inner edge. In these cases,
the flow is expected to pass through the inner sonic point in a time-dependent
manner and some sort of quasi-periodic oscillations cannot be ruled out.

\subsection{Nucleosynthesis in Hot Flows}

\noindent {\it Case B.1:} This case is chosen with such a set of parameters
that a standing shock forms at $x_s=13.9$. A very low accretion rate is
chosen so that the Compton cooling is negligible and the  flow remains
very hot (Comptonization factor $F_{\mathrm{Compt}}=0.1$). We show the
results both for the shock-free branch (dashed)
and the shocked branch (solid) of the solution. Figure 5a shows 
the comparison of the temperatures and densities (in units of $10^{-20}$ gm cm$^{-3}$ to 
bring in the same plot).  Figure 5b shows the comparison of the 
radial velocities. This  behaviour is similar to that shown in Case A.2.
Because the temperature is suitable for photo-dissociation,
we chose a very small set of species in the network (only 21 species
up to $^{11}\!B$ are chosen). Figure 5c shows the comparison of the abundances of proton (p),
$^{4}\!He$ and neutron (n). In the absence of the shock, the breaking up
of $^4\!He$ into n and p takes place much closer to the black hole, while the shock
hastens it due to higher temperature and density. Although initially the flow starts 
with $Y_p = 0.7425$ and $^4\!He=0.2380$, at the end of the simulation, only proton
($Y_p \sim 0.8786$) and neutron ($Y_n \sim 0.1214$) remain and the
rest of the species become insignificant.

Figure 5d shows the comparison of the height-integrated nuclear energy release
(units are as Fig. 2d). As the flow leaves the Keplerian disk at $x_K=481.4$, 
the deuterium and $^9\!Be$  are burnt instantaneously at the cost of some energy from the 
disk. At the end of deuterium burning at around $x=200$, the  {\it rp} and 
proton capture processes (mainly via $^{11}\!B(p,\gamma)3 ^4\!He$ which releases significant energy) 
and neutron capture ($^3\!He(n,p)T$) take place, but further in, $^3\!He$ 
(via $^3\!He(\gamma,p)D$) first and $^4\!He$ (mainly via 
$^4\!He(\gamma,n)^3\!He$ and $^4\!He(\gamma,p)T$, $T(\gamma,n)D$) 
subsequently, are rapidly dissociated. As soon as the entire helium is burnt out,
the energy release becomes negligible. This is because there is nothing left 
other than free protons and neutrons and hence no more reactions take place 
and no energy is released or absorbed. The solid curve is for the branch 
with a shock and the very long dashed curve is for the shock-free branch.
Inclusion of an opacity factor (which reduces photo-dissociation)
shifts the burning towards the black hole. The disk is found to be
completely stable even in presence of nucleosynthesis.

\noindent {\it Case B.2:} As discussed in Sect. 2, in extreme hard states, a black hole
may accrete very little matter in the Keplerian component and very large amount of 
matter in the sub-Keplerian component. To simulate this we used B.1 parameters, but 
${\dot m}=4$. The resulting solution is found to be unstable when shocks
are present. In Fig. 5b, we superimposed velocity variation without
nuclear energy (same as with nuclear energy as far as Case B.1 is concerned)
and with nuclear energy. The dash-dotted curve next to the un-shocked 
branch and dashed curve next to the shocked branch  show the resulting deviation.
While the branch without shock still remains stable, the other branch is distinctly
unstable as the steady-state solution is sub-sonic at the inner edge. The only 
solution available must be non-steady with oscillations near the sonic point.

\noindent {\it Case B.3:} In this case, accretion rate is chosen to be even smaller
(${\dot m}=0.001$) and the polytropic index is chosen to be $5/3$. The maximum 
temperature reaches  $T_9^{\mathrm{max}}=47$.  After leaving the Keplerian flow, the 
temperature and velocity of the flow monotonically increases.
Because of excessive temperature, $D$ and $^3\!He$ are photo-dissociated  immediately
after the flow leaves the Keplerian  disk at $x_K=84.4$. At around $x=30$, all
$^4\!He$ is photo-dissociated exactly as in Case B.1. Subsequently, the flow 
contains only protons and neutrons and there is no more energy release from the 
nuclear reactions. This behaviour is clearly seen in Fig. 6. The notations are the  same 
as in the previous run. This ultra-hot case is found to be stable since the energy
release took place far away from the black hole where the matter was moving slowly
and therefore the rate ($Q_{\mathrm{nuc}}$) was not high compared to that due to
viscous dissipation (units are as Fig. 2d).

\begin {figure}
\vbox{
\vskip -4.5cm
\hskip 0.0cm
\centerline{
\psfig{figure=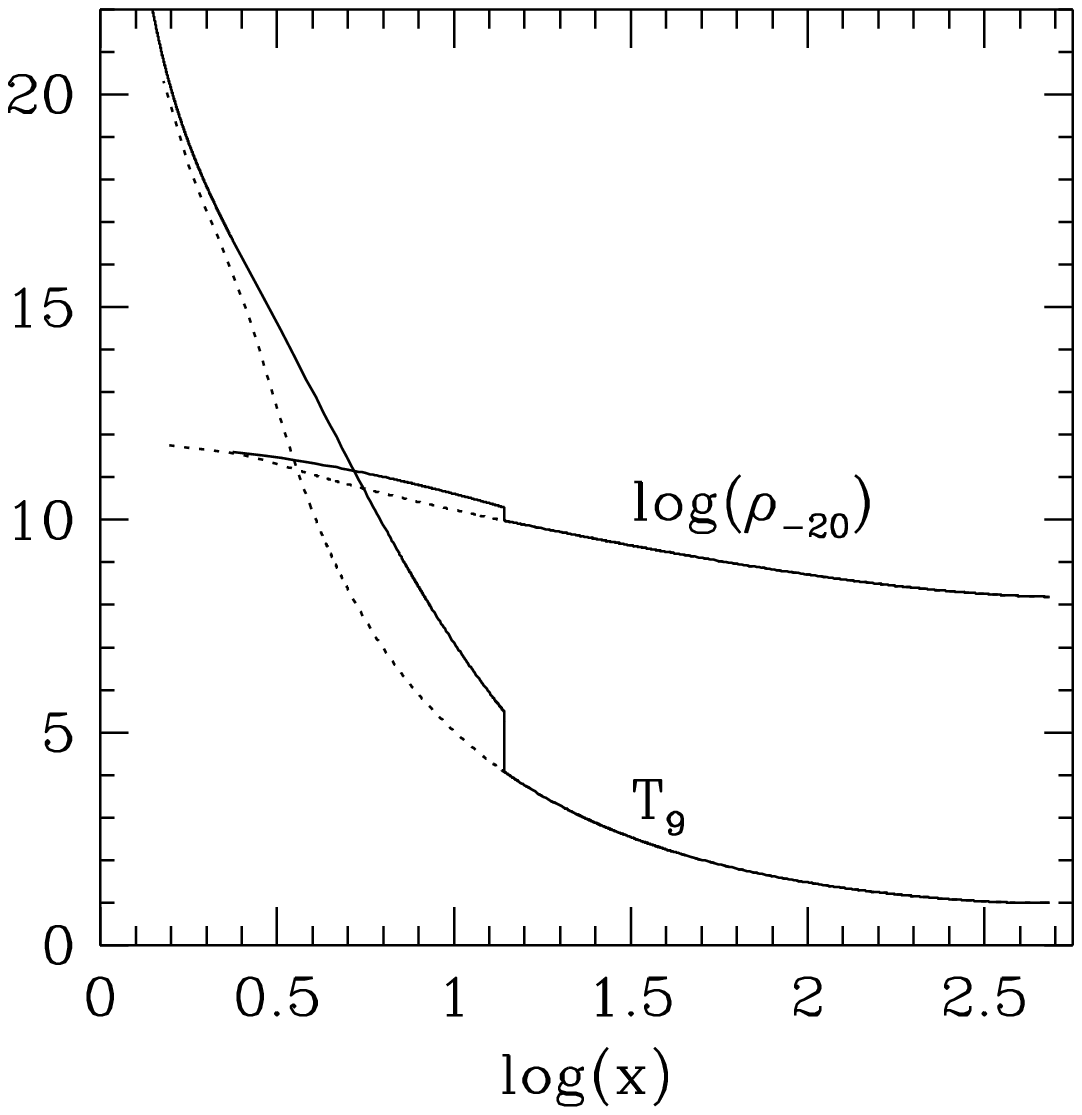,height=10truecm,width=10truecm,angle=0}}}
\vspace{-1.0cm}
\noindent{\small {\bf Fig. 5a}}
\end{figure}

\begin {figure}
\vbox{
\vskip -4.5cm
\hskip 0.0cm
\centerline{
\psfig{figure=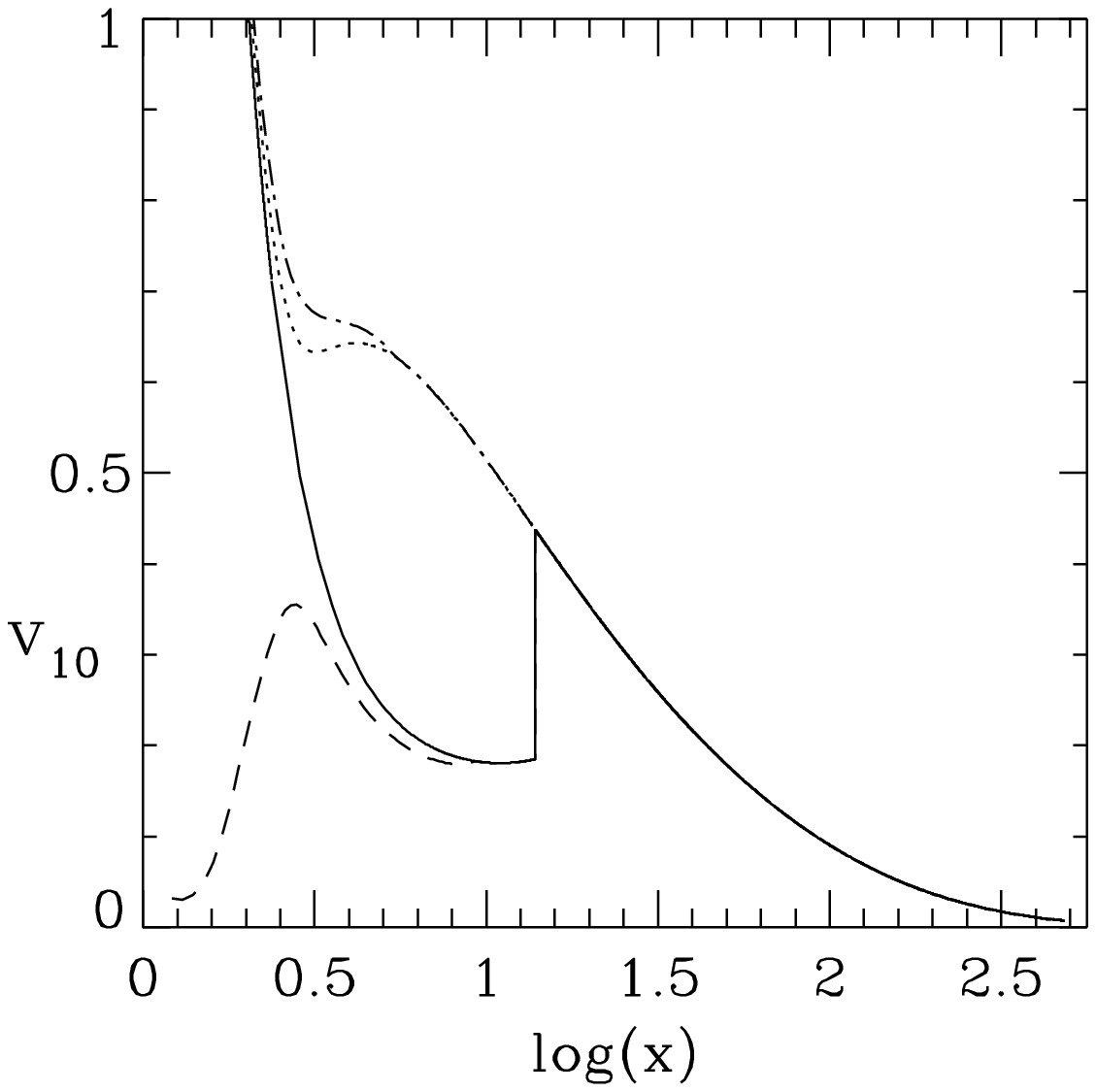,height=10truecm,width=10truecm,angle=0}}}
\vspace{-1.0cm}
\noindent{\small {\bf Fig. 5b}}
\end{figure}

\begin {figure}
\vbox{
\vskip -4.5cm
\hskip 0.0cm
\centerline{
\psfig{figure=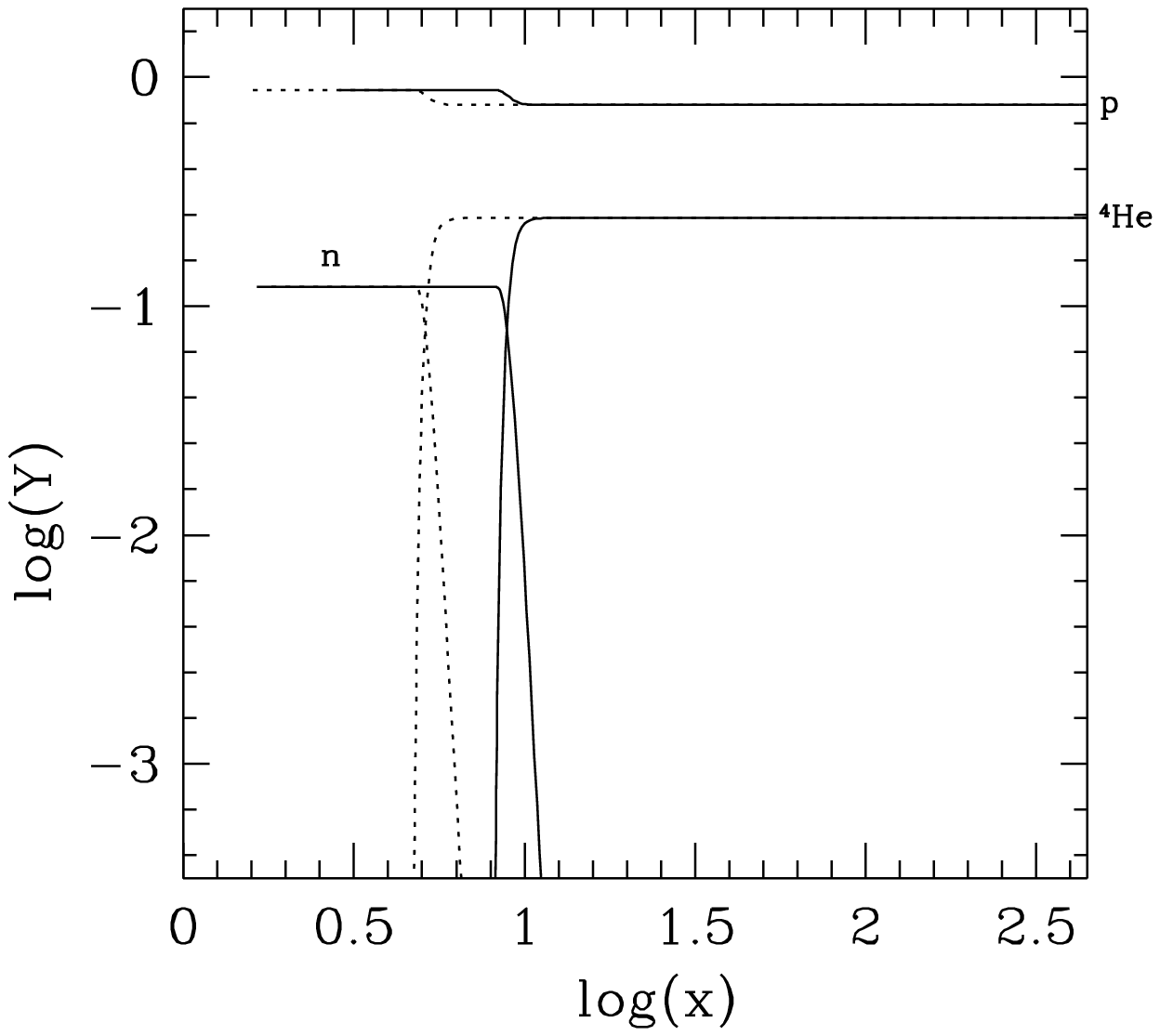,height=10truecm,width=10truecm,angle=0}}}
\vspace{-1.0cm}
\noindent{\small {\bf Fig. 5c}}
\end{figure}

\begin {figure}
\vbox{
\vskip -4.5cm
\hskip 0.0cm
\centerline{
\psfig{figure=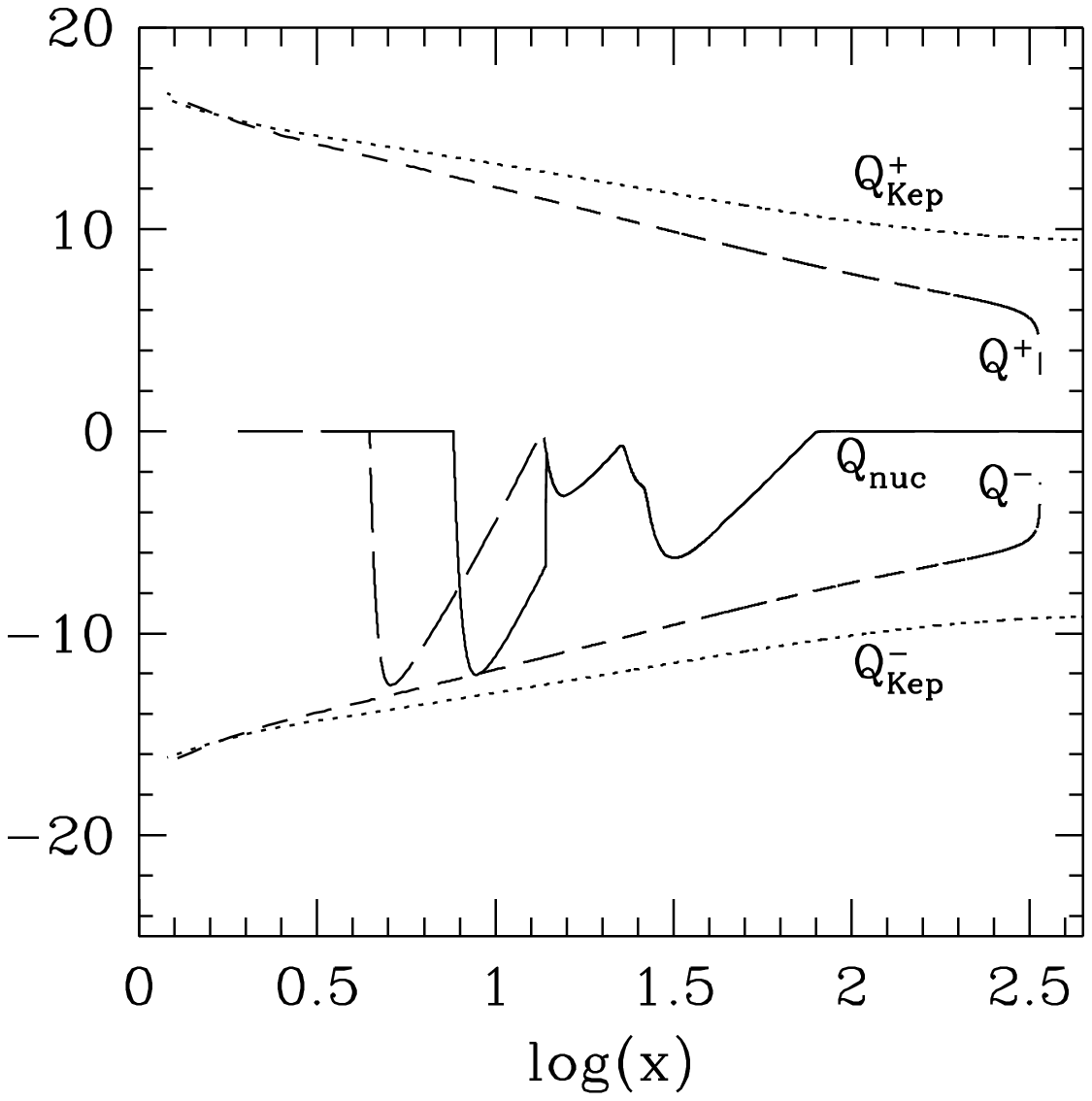,height=10truecm,width=10truecm,angle=0}}}
\vspace{-1.0cm}
\noindent{\small {\bf Fig. 5d}}
\end{figure}

\begin {figure}
\noindent {\small {\bf Fig. 5.} 
Variation of (a) ion temperature ($T_9$) and density ($\rho_{-20}$), (b) radial 
velocity $v_{10}$, (c) matter abundance $Y_i$ in logarithmic scale and (d) various forms of 
height-integrated specific energy release and absorption rates when the model parameters are $M=10M_\odot$, 
${\dot m}=0.01$, $\alpha_\Pi=0.05$
as functions of logarithmic radial distance ($x$ in units of Schwarzschild radius).
See text and Table 1 for other parameters of Case B.1 which is considered here.
Solutions in the stable branch with shocks are solid curves and those without the shock
are short dashed in (a-c). Curves in (d) are described in the text.
The ultra-hot temperature of the flow photo-dissociates $^4\!He$ into protons and neutrons.
The shocked branch (which is stable) causes such dissociation farther out from the
black hole than the unstable shock-free branch.
In (b), dot-dashed curve and dashed curves
are drawn when nuclear energy is taken into account and ${\dot m}=4$ is chosen (Case B.2).  }
\end{figure}

\begin {figure}
\vbox{
\vskip -2.0cm
\hskip 0.0cm
\centerline{
\psfig{figure=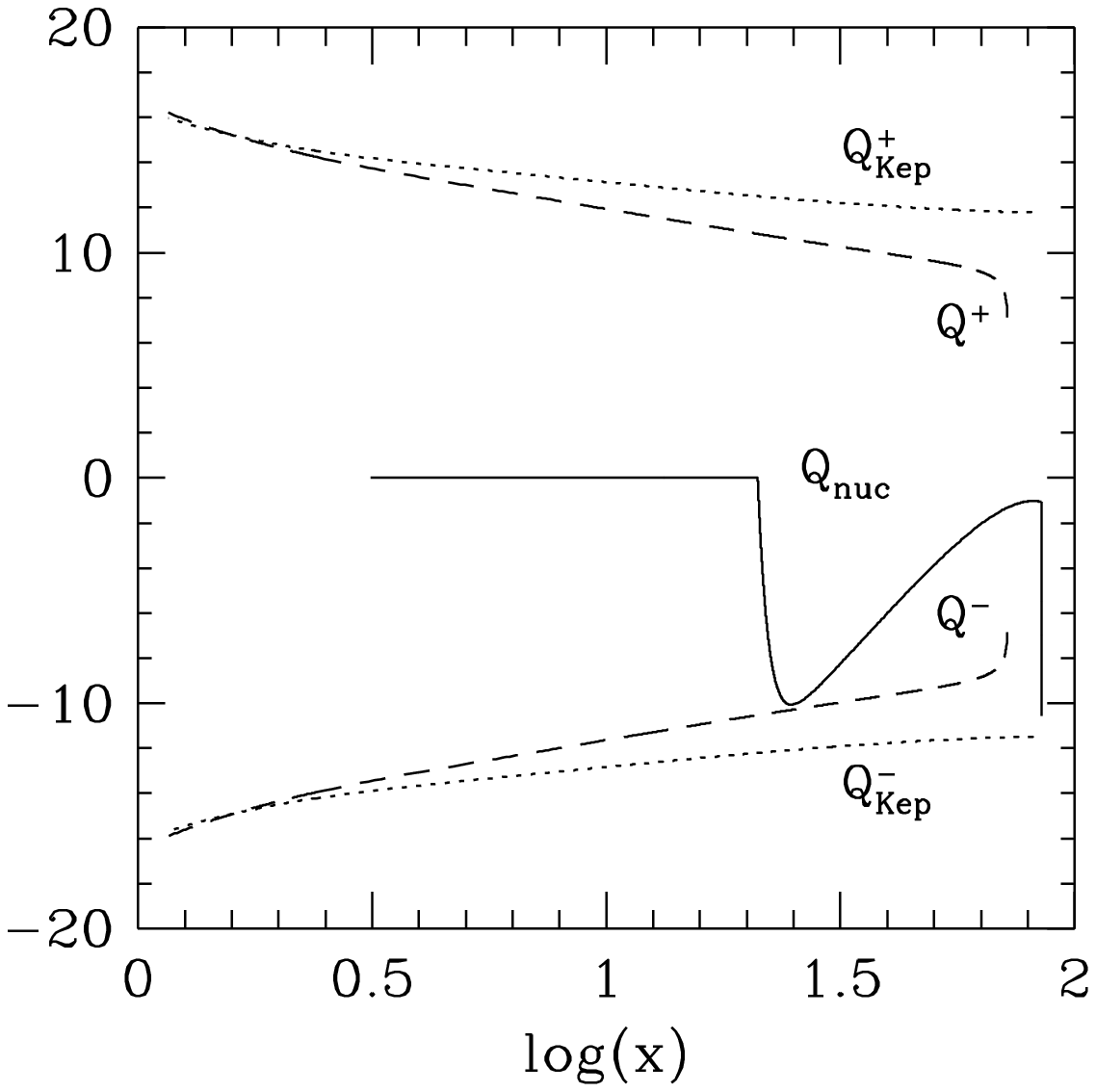,height=10truecm,width=10truecm,angle=0}}}
\vspace{0.5cm}
\noindent {\small {\bf Fig. 6.} Specific nuclear energy rate variation curve (solid) for a $\gamma=5/3$,
ultra-hot case ($T_9^{\mathrm{max}}=44$) as functions of logarithmic radial
distance ($x$ in units of Schwarzschild radius). The entire initial abundance is
photo-dissociated at $x\gsim 30$.  The viscous energy generation curve ($Q^+$) and
absorption curve ($Q^-$)  [both long dashed] are presented for comparison. $Q^\pm_{\mathrm{Kep}}$
(dotted) curves are the specific energy generation and absorption rates provided
the inner disks were Keplerian. $Q$s are in units of ergs cm$^{-2}$ sec$^{-1}$.
See Table 1 for parameters of Case B.3.}
\end{figure}

\noindent{\it Case B.4:} In this case, the net accretion rate is low
(${\dot m}=0.01$) but viscosity is high and the efficiency of emission 
is intermediate ($f=0.2$). That means that the temperature of the
flow is high ($F_{\mathrm{Compt}} = 0.1$, maximum temperature 
$T_9^{\mathrm{max}}=13$). Matter deviates 
from a Keplerian disk at around 
$x_K=8.4$. Assuming that the high viscosity is due to stochastic
magnetic field, protons would be drifted towards the black hole
due to magnetic viscosity, but the neutrons will not be 
drifted (Rees et al. 1982). They will generally circle around 
the black hole till they decay. This principle has been used 
to do the simulation in this case. The modified composition in one 
sweep is allowed to interact with freshly
accreting matter with the understanding that the accumulated neutrons
do not drift radially. After few iterations or sweeps the steady
distribution of the composition is achieved. Figure 7
shows the neutron distribution in the sub-Keplerian region. The formation
of a `neutron torus' is very apparent in this result.
In fact, the formation of a neutron disk is very generic in all the hot, highly viscous
accretion flows as also seen in Cases B.1-B.3 (for details, see, Paper 1).  
The nuclear reactions leading to the neutron torus
formation are exactly same as previous cases and are not described here.

\begin {figure}
\vbox{
\vskip -4.5cm
\hskip 0.0cm
\centerline{
\psfig{figure=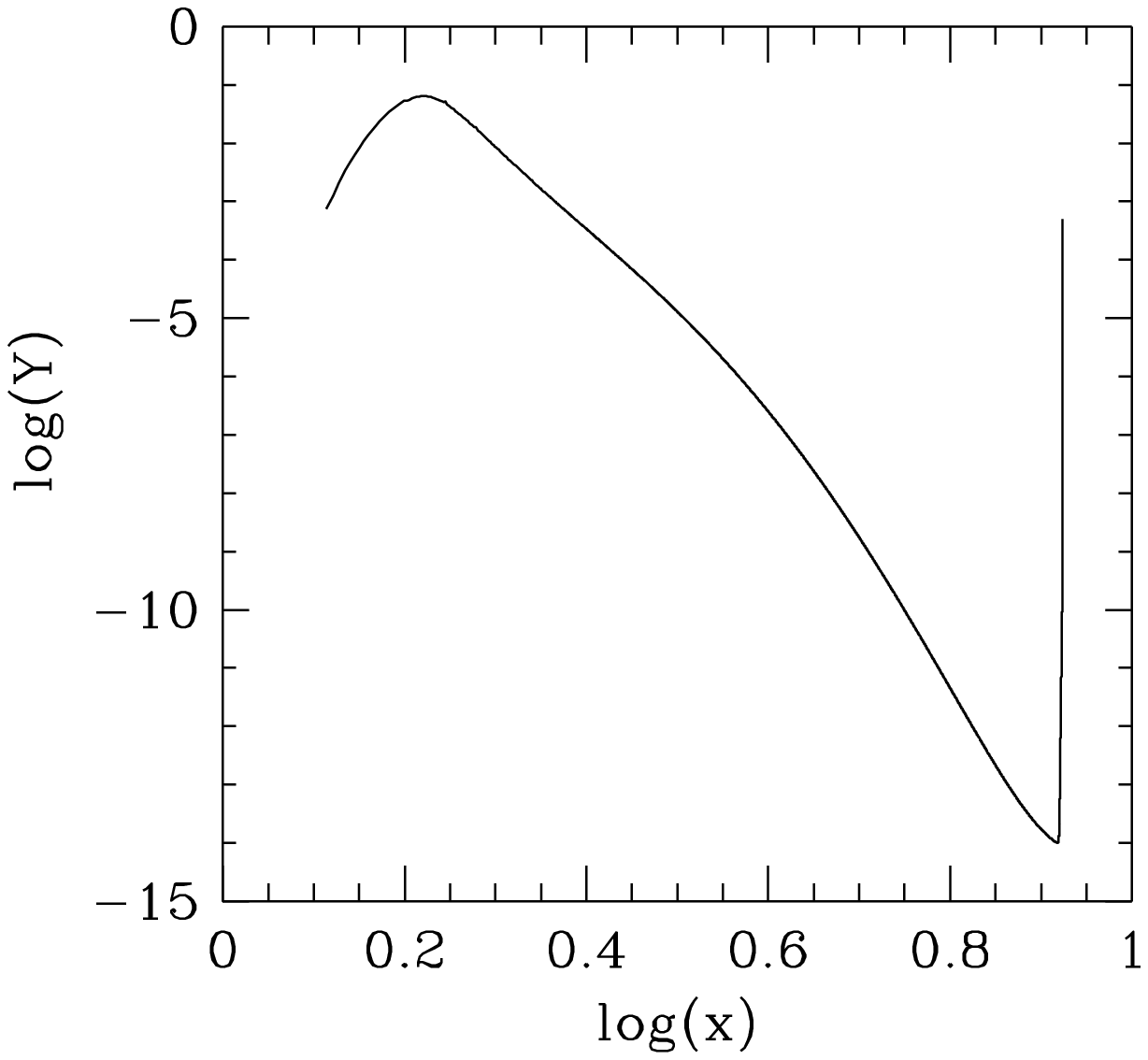,height=10truecm,width=10truecm,angle=0}}}
\vspace{-0.5cm}
\noindent{\small {\bf Fig. 7}}
\end{figure}

\begin {figure}
\noindent {\small {\bf Fig. 7.} Formation of a {\it neutron torus} in a hot inflow.
 Neutron abundance as a function of the logarithmic radial distance ($x$ in units of Schwarzschild radius).
See, Table 1 for parameters of Case B.4.}
\end{figure}

\subsection{Nucleosynthesis in Cooler Flows}

\noindent{\it Case C.1:} Here we choose a high-viscosity flow with a very high
accretion rate. Matter deviates from the Keplerian disk very close to the
black hole $x_K=4.8$. The flow in the centrifugal barrier is cooler (temperature
maximum $T_9^{\mathrm{max}}=0.8$). 
Here clearly, high viscosity
removes the centrifugal barrier completely and matter falls in almost freely.  
Due to very short residence time, no significant change in the composition 
takes place. Only a small amount of proton capture (mainly due 
to $^{11}\!B(p,\gamma)3 ^4\!He $, $^{16}\!O(p,\alpha)^{13}\!N$,
$^{15}\!N (p,\alpha) ^{12}\!C$, $^{18}\!O(p,\alpha)^{15}\!N $,
$^{19}\!F (p,\alpha)^{16}\!O $) takes place. A small amount of deuterium
dissociation also take place, but it does not change the energetics significantly. 
The flow is not found to be unstable in this case.

\noindent{\it Case C.2:} This is a test case for the proto-galactic accretion flow. In the
early phase of galaxy formation, the supply of matter is high, and the temperature of the
flow is very low. The viscosity may or may not be very high, but we choose very low
(presumably, radiative) viscosity ($\alpha = 10^{-4}$). The motivation is to use similar
parameters as were used in JAC while studying the nucleosynthesis in thick accretion disks.
The central mass $M=10^6 M_\odot$, the maximum temperature is $T_9^{\mathrm{max}} \sim 0.2$ 
and the Comptonization factor $F_{Compt}=0.001$. The temperature variation 
is similar to Fig. 2a when scaled down by a factor of $30$ (basically by the ratio of 
the $F_{\mathrm{Compt}}$ values). The velocity variation is similar to Fig. 2b  and is not repeated here.
Due to the low temperature, there is no significant change in the nuclear abundance. Note that since
thick accretion disks are rotation dominated, the residence time was very long 
in CJA simulation and there {\it was} significant change in composition  
even at lower temperatures. But in this case the flow radial velocity 
is very high and the residence time is shorter. The nuclear energy release is 
negligible throughout and is not shown.

\section{Nucleosynthesis Induced Instability}

CJA, while studying nucleosynthesis in cooler, mainly rotating disks, suggested that
as long as the nuclear energy release is smaller than the gravitational energy
release, the disk would be stable. In the present paper, we find that this 
suggestion is still valid. Indeed, even when momentarily the nuclear energy
release or absorption is as high as the gravitational energy release (through
viscous dissipation), the disk may be stable. 
For instance, in case A.1 (Fig. 2d) at around $x=4$ these rates are similar. Yet the velocity,
temperature and density distributions (Fig. 2a-b) remain unchanged. In  Case A.3, 
$Q_{\mathrm{nuc}}$ is several magnitudes greater than viscous energy release $Q^+$
and the thermodynamic quantities are indeed disturbed to the extent that the
flow with same injected quantities (with the same density and velocity
and their gradients) at the outer edge does not become supersonic at the inner edge.
In these cases, the flow must be unsteady in an effort to search for the
`right' sonic point to enter into the black hole. On the other hand, 
ultra-hot cases like B.2 show deviation in non-shocked solution while the
shocked solution is unstable. 

The general behaviour suggests that the present model of accretion disks is more
stable under nuclear reactions compared to the earlier, predominantly rotating model. Here, the radial velocity
($v$) spreads energy release or absorption radially to a distance $v\tau_D(\rho,T)=v N_D/{\dot N}_D$ cm, where,
$N_D$ is the number density of, say, Deuterium and ${\dot N}_D$ is its depletion
rate. For a free fall, $v\sim x^{-1/2}$, while for most nuclear reactions, 
$\tau_D (\rho,T) \sim x^n$, with $n>>1$ (since reaction rates are strongly 
dependent on density and temperature). Thus, $Q_{\mathrm{nuc}}$ for the destruction 
of a given element spreads out farther away from the black hole,
but steepens closer to it. Large $dQ_{{nuc}}/dx$ causes instability since the derivatives such as $dv/dx$
at the inner regions (including the sonic point) become imaginary.

\section{Discussions and Conclusions}

In this paper, we have explored the possibility of nuclear reactions in
inner  accretion flows. Because of high radial motion and ion pressure, matter
deviates from a Keplerian disk close to the black hole. The temperature
in this region is controlled by the efficiencies of bremsstrahlung and 
Comptonization processes (CT96, C97) and possible heating by magnetic fields
(Shapiro 1973): for a higher
Keplerian rate and higher viscosity, the inner edge of the Keplerian
component comes closer to the black hole and the sub-Keplerian  region becomes
cooler (CT95). The nucleosynthesis in this soft state of the
black hole is quite negligible. However, as the viscosity is decreased to
around $0.05$ or less, the inner edge of the Keplerian component moves away and the 
Compton cooling becomes less efficient due to the paucity of the supply of soft photons.
The sub-Keplerian region, though cooler by a factor of about 
$F_{\mathrm{Compt}}=0.01$ to $0.03$ from that of the value obtained 
through purely hydrodynamical calculations of C96, is still high enough to 
cause significant nuclear reactions to modify compositions.
The composition changes very close to the black hole, 
especially in the centrifugal-pressure-supported denser region, 
where matter is hotter and slower.

The degree of change in compositions which takes place in the Group A and B calculations,
is very interesting and its importance must not be underestimated. 
Since the centrifugal-pressure-supported region can be treated as an effective
surface of the black hole which may generate winds and outflows in the
same way as the stellar surface (Chakrabarti 1998a,b; Das \& Chakrabarti 1999), 
one could envisage that the winds produced in this region would
carry away a modified composition and contaminate the atmosphere of the
surrounding stars and the galaxy in general. 

One could estimate the contamination of the galactic metalicity
due to nuclear reactions. For instance, in Case A.1, $^{12}\!C$, $^{16}\!O$, 
$^{20}\!Ne$, $^{30}\!Si$, $^{44}\!Ca$ and  $^{52}\!Cr$
are found to be over-abundant in some region of the disk. Assume that, on an average,
all the $N$ stellar black holes are of equal mass $M$ and have a non-dimensional
accretion rate of around ${\dot m} \sim 1$ (${\dot m}={\dot M}/{\dot M}_{\mathrm{Edd}}$).
Let  $\Delta Y_i$ (few times $10^{-3}$) be the typical 
change in composition of this matter during the run and let $f_w$ be the fraction of 
the incoming flow that goes out as winds and outflows (could be from 
ten percent to more than a hundred percent when disk evacuation occurs), then 
in the lifetime of a galaxy (say, $10^{10}$yrs), the total `change'
in abundance of a particular species deposited in the surroundings by all the stellar
black holes is given by:
$$
<\Delta Y_i>_{\mathrm{small}} \cong 10^{-7}(\frac{\dot m}{1})(\frac{N}{10^6}) 
(\frac{\Delta Y_i}{10^{-3}})(\frac{f_w}{0.1}) (\frac {M}{10 M_\odot}) 
(\frac{T_{\mathrm{gal}}}{10^{10}} Yr) (\frac{M_{\mathrm{gal}}}{10^{11} M_\odot})^{-1}.
\eqno{(2)}
$$
The subscript `small' is used here to represent the contribution from small black holes.
We also assume a conservative estimate that there are $10^6$ such stellar black holes
in a galaxy, the mass of the host galaxy is around $10^{11}M_\odot$ and the lifetime of
the galaxy during which such reactions are going on is about $10^{10}$Yrs. 
We also assume that $\Delta Y_i \sim 10^{-3}$ and a fraction of ten percent of matter is blown off as winds.
The resulting $<\Delta Y_i> \sim 10^{-7}$ may not be very significant if one considers
averaging over the whole galaxy. However, for a lighter galaxy $<\Delta Y_i>$ 
could be much higher. For example, for $M_{gal}=10^{9}M_\odot$, $<\Delta Y_i> \sim 10^{-5}$.
This would significantly change the average abundances of $^{30}\!Si$, $^{44}\!Ca$ and  $^{52}\!Cr$.
On the other hand, if one concentrates on the region of the outflows only,  
the change in abundance is the same as in the disk, and should be detectable (e.g., through
line emissions). One such observation of stronger iron-line emission 
was reported for SS433 (Lamb et al. 1983; see also Arnould \& Takahashi 1999, 
for a recent discussion on galactic contaminations).

When we consider a case like A.3, we find that
$^{12}\!C$, $^{16}\!O$, $^{20}\!Ne$, and $^{28}\!Si$ are increased by about $10^{-3}$ in some regions. In this case, the average
change of abundance due to accretion onto the massive black hole situated at the galactic centre would be,
$$
<\Delta Y_i>_{\mathrm{big}} \cong few \times 10^{-8}(\frac{\dot m}{1})
(\frac{\Delta Y_i}{10^{-3}}) (\frac{f_w}{0.1}) (\frac {M}{10^6 M_\odot})
(\frac{T_{\mathrm{gal}}}{10^{10}} Yr) (\frac{M_{\mathrm{gal}}}{10^{11} M_\odot})^{-1}.
\eqno{(3)}
$$
Here, we have put `big' as the subscript to indicate the contribution from the massive black hole.
Even for a lighter galaxy, e.g., of mass $M_{\mathrm{gal}}=10^9 M_\odot$, $\Delta Y_i = 10^{-6}$ which
may not be significant. If one considers only the regions of outflows, contamination may not be negligible.

A few related questions have been asked lately: Can lithium be 
produced in black hole accretion? We believe not. 
The spalletion reactions (Jin 1990; Yi \& Narayan 1997) which may produce such elements
assuming that a helium beam hits a helium target in a disk. Using
a full network, rather than only He-He reaction, we find that the
hotter disks where spalletion would have been important also photo-dissociate
(particularly due to the presence of photons from the Keplerian disk)
helium to deuterium and then to protons and neutrons before any significant
lithium could be produced. Even when photo-dissociation is very low (when the
Keplerian disk is far away, for instance), or when late-type  stellar
composition is taken as the initial composition, we find that the $^7\!Li$ production
is insignificant, particularly if one considers more massive black holes ($M\sim 10^8 M_\odot$).

Recently, it has been reported by several authors (Martin et al. 1992; 1994; Fillipenko et al. 1995;  Harlaftis et al. 1996) 
that a high abundance of $Li$ is observed in late type stars which are also companions
of black hole and neutron star candidates. This is indeed surprising since
the theory of stellar evolution predicts that 
these stars should have at least a factor of ten lower  $Li$ abundance.
These workers have suggested that this excess $Li$ could be produced in 
the hot accretion disks. However, in Paper 1 as well as in our Cases   A and B 
computations we showed that $Li$ is not likely to be produced in accretion disks.
Indeed, we ran several cases with a mass fraction of He as high as 0.5 to 0.98,
but we are still unable to produce $Li$ with a mass fraction more than $10^{-10}$.
Recent work of Guessoum \& Kazanas (1999) agrees with our conclusion that
profuse neutrons would be produced in the disk. They farther suggested that these
energetic neutrons can produce adequate $Li$ through spalletion 
reactions with the $C$, $N$, and $O$ that is present in the atmospheres of these stars.
For instance, in Cases B.1 and B.3 we see that neutrons could have an abundance of 
$\sim 0.1$ in the disk. Since the production rate is similar to what 
Guessoum \& Kazanas (1999) found, $Li$ should also be produced on 
stellar surface at a similar rate.

What would be the neutrino flux on earth if nucleosynthesis does take place?
The energy release by neutrinos (the pair neutrino process, the photoneutrino process
and the plasma neutrino process) can be calculated using the prescription of Beaudet et al. (1967, hereafter BPS; see also Itoh et al. 1996) provided 
the pairs are in equilibrium with the radiation field. However, in the case
of accretion disks, the situation is significantly different from that inside a star (where
matter is in static equilibrium). Because of rapid infall, matter density is much lower
and the infall time scale could be much shorter compared to the time-scale of various neutrino
processes, especially the pair and photo-neutrino processes. As a result, the pair density need not attain
equilibrium. One important thing in this context is the 
opacity ($\tau_{\mathrm{pair}}$) of the pair process. Following treatments of Colpi et al. (1984) 
we find that $\tau_{\mathrm{pair}} <1$ for all our cases, and therefore
pair process is expected to be negligible (for Case B.2, $\tau_{\mathrm{pair}}$ is the highest [$0.9$]).
Park (1990a,b), while studying pair creation processes in spherical accretion,
shows that even in the most favourable condition, the ratio of positron ($n_+$) and ion ($n_i$)
is no more than $0.05$. A simple analysis suggests that neutrino production rate is many orders of magnitude smaller compared to what the equilibrium solutions of BPS and Itoh et al. would predict.
Thus, we can safely ignore the neutrino luminosity.

When the nuclear energy release or absorption is comparable to the gravitational
energy release through viscous processes, we find that the disk is still stable.
Stability seems to depend on how steeply the energy is released or absorbed in the
disk. This in turn depends on $\tau_D v$, the distance traversed inside the disk
by the element contributing the highest change of energy before depleting
significantly. Thus, an ultra-hot case (Group B) can be stable even though
a hot (Group A) case can be unstable as we explicitly showed by 
including nuclear energy release.  In these `unstable' cases, we find that the
steady flow does not satisfy the inner boundary condition and becomes subsonic
close to the horizon. This implies that in these cases the flow must become 
non-steady, constantly searching for the supersonic branch to 
enter into the black hole. This can induce oscillations as have been 
found elsewhere (Ryu et al. 1997). In such cases, one is required to do
time dependent simulations (e.g., Molteni et al.  1994, 1996) to include nuclear reactions. This will be
attempted in future.

We thank the referee for many helpful comments. This research is
partially supported by DST grant under the project ``Analytical
and numerical studies of astrophysical flows around black holes and neutron stars"
with SKC.


{}

\end{document}